\documentclass{aa}  
\usepackage{graphicx,url}
\usepackage[normalem]{ulem}
\usepackage{mathrsfs,amssymb,amsmath}
\usepackage[usenames]{color}
\usepackage{subfigure}
\usepackage[varg]{txfonts}
\usepackage{booktabs}
\usepackage{multirow}
\usepackage[breaklinks, colorlinks, citecolor=blue]{hyperref}
\newcommand{\Vr}{\mathbf{r}}
\newcommand{\ud}{{\rm d}}

\newcommand{\bperp}{B_\perp}
\begin{document}

   \title{First look at the multiphase interstellar medium with synthetic observations of low-frequency Faraday tomography} 

   \titlerunning{Multiphase interstellar medium with synthetic observations of low-frequency Faraday tomography}
   \authorrunning{Bracco et al.}

   \author{Andrea Bracco\inst{1,2} 
             \and
            Evangelia Ntormousi\inst{3,4}
            \and
            Vibor Jeli\'c\inst{1}
            \and
            Marco Padovani\inst{5} 
            \and
            Barbara \v{S}iljeg\inst{1,6,7}
            \and
            Ana Erceg\inst{1}
            \and\\
            Luka Turi\'c\inst{1}
            \and
            Lana Ceraj\inst{1}
            \and
            Iva \v{S}nidari\'c\inst{1}
          }

   \institute{Ru{\dj}er Bo\v{s}kovi\'c Institute, Bijeni\v{c}ka cesta 54, 
              10000 Zagreb, Croatia \\
              \email{abracco@irb.hr}
             \and
             Laboratoire AIM, CEA / CNRS / Universit\'e Paris-Saclay, 91191 Gif-sur-Yvette, France
             \and
             Scuola Normale Superiore,
            Piazza dei Cavalieri, 7
            56126 Pisa, Italy
            \and
            Institute of Astrophysics, Foundation for Research and Technology-Hellas, Vasilika Vouton, GR-70013 Heraklion, Greece
            \and
            INAF--Osservatorio Astrofisico di Arcetri, 
              Largo E. Fermi 5, 50125 Firenze, Italy
            \and
            ASTRON, the Netherlands Institute for Radio Astronomy, Oude Hoogeveensedijk 4,7991 PD Dwingeloo, The Netherlands
            \and
            Kapteyn Astronomical Institute, University of Groningen, P.O. Box 800, 9700 AV, Groningen, The Netherlands\\
             }

   \date{Received: October 15, 2021; accepted: April 15, 2022}

 
  \abstract
  {Faraday tomography of radio polarimetric data below 200 MHz from the LOw Frequency ARray (LOFAR) are providing us with a new perspective on the diffuse and magnetized interstellar medium (ISM).
  Of particular interest is the unexpected discovery of Faraday-rotated synchrotron polarization associated with structures of neutral gas, as traced by atomic hydrogen (HI) and dust.\\
  Here we present the first in-depth numerical study of these LOFAR results. We produce and analyze comprehensive synthetic observations of low-frequency synchrotron polarization from magneto-hydrodynamical (MHD) simulations of colliding super shells in the multiphase ISM, already presented in  \citet{Ntormousi2017}. \\
  Using an analytical approach to derive the ionization state of the multiphase gas, we define five distinct gas phases over more than four orders of magnitude in gas temperature and density, ranging from hot, and warm, fully ionized gas to cold neutral medium.  \\
  We focus on establishing the contribution of each gas phase to synthetic observations of both rotation-measure and synchrotron polarized intensity below 200 MHz. We also investigate the link between the latter and synthetic observations of optically thin HI gas. \\
  We find that, not only the fully ionized gas but also the warm partially ionized and neutral phases strongly contribute to the total rotation measure and polarized intensity. However, the contribution of each phase to the observables strongly depends on the choice of integration axis and the orientation of the mean magnetic field with respect to the shell collision axis.
  Strong correlation between HI synthetic data and synchrotron polarized intensity, reminiscent of LOFAR results, is obtained with lines of sight perpendicular to the mean magnetic field direction.   \\
  Our study suggests that multiphase modelling of MHD processes is needed in order to interpret observations of the radio sky at low frequency. This work is a first step toward understanding the complexity of low-frequency synchrotron emission that will be soon revolutionized by large-scale surveys with LOFAR and the Square Kilometre Array.  
  }
 
   \keywords{ISM: magnetic fields; ISM: structure; ISM: bubbles; methods: numerical; polarization; radio continuum: ISM}

   \maketitle
%

\section{Introduction}\label{sec:intro}

Magnetic fields are fundamental ingredients of the turbulent cascade that steers and shapes the diffuse interstellar gas from kiloparsec to sub-parsec scales, where star formation occurs \citep[see e.g. review by][]{Hennebelle2019}. 
However, the interaction of magnetic fields with interstellar matter is very hard to characterize observationally. 
One reason for this difficulty is that this interaction is not only multi-scale, but also multiphase.
Depending on the thermodynamics of interstellar gas a number of distinct phases can be identified based on observations \citep{Heiles2012,Ferriere2020}. Fully ionized gas is at temperatures above $10^6$ K (Hot Ionized Medium, HIM) or at $\sim 10^4$ K (Warm Ionized Medium, WIM), based on X-ray, UV, and optical spectroscopy \citep[e.g.,][]{Snowden1997,Jenkins2013,Krishnarao2017}. UV spectroscopy of the local ISM has also suggested the presence of gas at lower temperatures ($\sim 5000$ K) with ionization fraction of about 0.5 \citep[Warm Partially Ionized Medium, WPIM,][]{Fitzpatrick1997,Redfield2004}. The mostly neutral phases in the diffuse ISM (with ionization fractions below $10^{-2}$) are well-known through line emission of atomic hydrogen (HI) at 21 cm. The HI gas is a mixture of bi-stable gas composed of warm neutral medium (WNM), at temperatures of $\sim$8000 K, and a cold neutral medium (CNM), with corresponding temperature of $\sim$50 K \citep{Field1965,Wolfire2003}. 
Being subject to thermal instability, HI gas also contains an unstable, lukewarm, neutral medium (LNM), which can be considered an intermediate phase between the two stable phases \citep[see Table~\ref{tab:phases} for a summary of the phases; see also][]{Saury2014, Marchal2019}.

Synchrotron emission and polarization are the main observational probes of interstellar magnetic fields \citep{Haslam1982,Reich1986,Davies1996,Guzman2011,Mozdzen2017,Mozdzen2019,Beck2019}. Therefore, if we want to understand magneto-hydrodynamical (MHD) turbulence in the ISM, we need to correctly interpret synchrotron data. Polarimetric observations of the LOw Frequency ARray \citep[LOFAR,][]{vanHaarlem2013} below 200 MHz recently started questioning our understanding of how synchrotron emission propagates throughout the diffuse and magnetized ISM. 
In particular, diffuse synchrotron emission is not expected to be specifically related to any gas phase listed above. It is the result of the interaction of cosmic ray electrons (CRe) and magnetic fields that are ubiquitous in the diffuse ISM \citep{PadovaniGalli2018}. However, in a number of studies, LOFAR observations revealed a striking morphological correlation between the structure of the observed synchrotron polarization and structures of neutral ISM, both traced by HI emission \citep{Kalberla2016,Jelic2018,Bracco2020b,Turic2021} and interstellar dust \citep{Zaroubi2015,vanEck2017,Turic2021}. 

Below 1~GHz, Faraday rotation complicates the interpretation of these observations \citep[i.e.,][]{Beck2015}. Magnetic fields and thermal electrons in the ionized multiphase gas along the line of sight (LOS) Faraday rotate the diffuse synchrotron polarized emission. The observed link between LOFAR polarization and neutral phases must be related to the full complexity of the magneto-ionic ISM, where synchrotron emission and Faraday rotation are mixed. 
A powerful technique used to disentangle various contributions of magneto-ionic medium along the LOS is called Faraday tomography \citep{Burn1966,Brentjens2005}. This technique takes radio-polarimetric data and decomposes the observed polarized synchrotron emission by the amount of Faraday rotation it experiences along the LOS. Faraday tomography maps the 3D relative distribution of the intervening magneto-ionic ISM based on Faraday depth. This quantity represents the specific amount of rotation measure along the LOS, which is the integrated effect of magnetic fields and thermal-electron density.

In light of Faraday tomography, which is sensitive to ionized gas, the correlation with the neutral phases revealed by LOFAR data is even more interesting. The question that arises is whether LOFAR is able to detect small amounts of Faraday depth coming from neutral clouds \citep[as discussed in][]{Bracco2020b} or if LOFAR is directly sensitive to synchrotron polarization associated to WNM, LNM, and CNM \citep[as first suggested by][]{vanEck2017}. This last hypothesis would imply that Faraday rotation in the ionized gas fully depolarizes synchrotron emission in the WIM and in the HIM, highlighting synchrotron polarization from the neutral phases. Any of the two scenarios suggests that LOFAR is providing us with a completely new perspective on the diffuse ISM.  

In order to investigate in depth these observations and study the complex, non-linear dependencies of synchrotron emission with the multiphase and magnetized ISM, a thorough analysis of MHD numerical simulations is needed. Synthetic observations of Faraday tomography from MHD numerical simulations have been already presented in recent works \citep{Basu2019,Seta2021}. However, to our knowledge, the multiphase aspect of the problem has never been addressed before. 

Hence we present the first synthetic low-frequency radio polarimetric observations of MHD simulations of a multiphase ISM. Since shells and loops are typical features observed in synchrotron emission \citep[e.g.,][]{berkhuijsen1971a,vidal15,Panopoulou2021,Erceg2022}, we have chosen to analyze synthetic observations of Faraday tomography at LOFAR frequencies from simulations of two colliding super shells presented in \citet{Ntormousi2017}. Our effort is only a first step in understanding the diffuse radio emission at low frequencies as a function of the ionization state of the ISM. A better knowledge of the diffuse synchrotron emission of the Galaxy will be crucial for interpreting upcoming large-scale surveys from LOFAR \citep[e.g. the LOFAR two-meter Sky Survey - LoTSS,][]{Shimwell2017} and the Square Kilometre Array in the future \citep{Dewdney2009}.

The paper is organized as follows. In Sect.~\ref{sec:methods} we describe the methodology used to model synchrotron emission and polarization below 200 MHz. We also present the MHD simulations and detail how we estimated the ionization state of the multiphase gas. Section~\ref{sec:results} presents our main results, which include: maps of rotation measure (Sect.~\ref{ssec:rmmaps}); the distinct contribution of magnetic fields and electrons to the rotation measure (Sect.~\ref{ssec:rmneb}); the correlation of the multiphase gas both with rotation measure (Sect.~\ref{ssec:phaserm}) and with polarized intensity based on Faraday tomography (Sect.~\ref{ssec:tomo} and Sect.~\ref{ssec:phasepi}). These results are discussed in Sect.~\ref{sec:discussion}, while  Sect.~\ref{sec:sum} summarizes and concludes the paper. The manuscript has two appendices.  

\section{Methods}\label{sec:methods}
In this section we describe the methodology and the formalism to model intrinsic\footnote{The term "intrinsic" refers to the synchrotron emission at low radio frequencies without the effect of Faraday rotation.} synchrotron emission (Sect.~\ref{ssec:emissivity}), Faraday rotation and synthetic Faraday cubes (based on Faraday tomography, Sect.~\ref{ssec:RM}) from the MHD simulations presented in Sect.~\ref{ssec:sims}. For more details on Faraday tomography, please refer to \citet{Burn1966}, \citet{Brentjens2005}, and \citet{Ferriere2020}.

We model the total synchrotron emission at frequency $\nu$ by producing synthetic observations of Stokes $I_{\nu}$, while we model the corresponding linear polarization by synthetic observations of Stokes $Q_{\nu}$ and $U_{\nu}$. Because of the Faraday rotation angle’s proportionality to the $\lambda^2$, modelling it accurately is crucial for observations below 200 MHz. 
This would not be necessary for models of synchrotron emission at higher radio frequencies (> 10 GHz). 

The methodology described in Sect.~\ref{ssec:emissivity} and Sect.~\ref{ssec:RM} is general and can be applied to any MHD simulation that provides magnetic field ($\vec{B}= b_x \hat{x}+b_y \hat{y}+b_z \hat{z}$\footnote{Here ($\hat{x},\hat{y},\hat{z}$) are the normal vectors of the orthonormal base that defines the simulated data cubes.}) and an estimate of the number density of thermal electrons, $n_e$, in 3D. 




\subsection{Intrinsic synchrotron emissivity}\label{ssec:emissivity}

We model intrinsic synchrotron total and polarized emission following \citet[][hereafter P21]{Padovani2021}. As CRe propagate through the ISM, they lose energy by a number of mechanisms that involve interactions
with matter, magnetic fields, and radiation \citep{Longair2011}. These processes deplete the population of CRe
and change their original energy spectrum, $j_e(E)$\footnote{I.e. number of electrons per unit energy, time, area, and solid angle.}, where $E$ is the energy of the CRe. A correct model of $j_e(E)$ is important as it determines the amount of specific emissivity of intrinsic synchrotron emission at frequency $\nu$. The specific emissivity can be split into two components linearly polarized along and across the component of the magnetic field perpendicular to the LOS, $\vec{B}_{\perp}$, as follows
\begin{eqnarray}\label{eq:epsnu}
\varepsilon_{\nu,\|}(\Vr) &=& \int_{m_{e}c^{2}}^{\infty}\frac{j_{e}(E)}{v_{e}(E)}P_{\nu,\|}^{\rm em}(E,{\bperp}(\Vr))\,\ud E,\\\nonumber
\varepsilon_{\nu,\perp}(\Vr) &=& \int_{m_{e}c^{2}}^{\infty}\frac{j_{e}(E)}{v_{e}(E)}P_{\nu,\perp}^{\rm em}(E,{\bperp}(\Vr))\,\ud E.
\end{eqnarray}
In Eq.~(\ref{eq:epsnu}), $v_e$ is the electron velocity, $m_e$ is the electron mass, $c$ is the speed of light, $P^{\rm em}_{\nu, \perp~\rm{\rm or}~\parallel}$ are the power per unit frequency emitted by an electron of energy $E$ at frequency $\nu$  for the two polarizations, and $B_{\perp}$ is the strength of $\vec{B}_{\perp}$ at position $\vec{r}$. For more details on Eq.~(\ref{eq:epsnu}) we refer the reader to P21, \citet{GinzburgSyrovatskii1964}, and \citet{Rybicki1979}.

The main difference of the P21 approach compared to previous works is that it includes realistic observational constraints on $j_e(E)$, set by considering the energy dependence of the spectral energy slope \citep[e.g.,][]{Sun2008,Waelkens2009,Reissl2019,Wang2020}. Following P21, in our models we consider a uniform spatial distribution of CRe and we use the $j_e(E)$ from \citet{Orlando18}. This CRe energy spectrum is based on multifrequency observations, from radio to $\gamma$-rays, and Voyager-1 measurements, and is representative of most of the local radio synchrotron emission within $\sim$1 kpc from the Sun. The use of a data-driven dependence of $j_e(E)$ with $E$, as discussed in P21, is particularly relevant at low radio frequencies. Standard approaches that consider a single power-law slope, of the kind $j_e \propto E^{s}$ with $s = -2$ or $-3$ depending on the energy range of the CRe \citep[e.g.,][]{Sun2008,Waelkens2009,Wang2020}, strongly bias the estimate of the synchrotron emissivities in the diffuse ISM toward flatter synchrotron spectral energy distributions (see P21 for more details).

We build synthetic maps of the total synchrotron emission, Stokes $I_{\nu}$, by integrating the quantity $\varepsilon_{\nu,\|}(\Vr) + \varepsilon_{\nu,\perp}(\Vr)$ along any given LOS of the simulated cubes.

\subsection{Faraday rotation and synthetic Faraday cubes}\label{ssec:RM}

In polarization, the derivation of the Stokes $Q_{\nu}$ and $U_{\nu}$ maps is more complicated in the presence of Faraday rotation. In this work we consider the case in which Faraday rotation is fully mixed with synchrotron emission, giving rise to differential Faraday rotation \citep[e.g.,][]{Sokoloff1998}. Each slice in the simulated cubes contributes both to synchrotron emission and Faraday rotation.  This means that the synthetic synchrotron Stokes $Q_{\nu}$ and $U_{\nu}$ are not only the result of integrating the corresponding emissivities along the LOS, as it was done in P21 neglecting the effect of Faraday rotation. Instead, we introduce effective synchrotron emissivities in polarization, $\tilde{\varepsilon}_{\nu,Q}$ and $\tilde{\varepsilon}_{\nu,U}$, by modifying eqs. (6) and (7) in P21 and defining the specific polarized emissivity at the $i$-th slice along a given LOS $\vec{r}$ as 
\begin{equation}\label{eq:emp}
    \varepsilon_{\nu,P} (\vec{r}_i)  = \varepsilon_{\nu,\perp} (\vec{r}_i) - \varepsilon_{\nu,\parallel} (\vec{r}_i).
\end{equation}
Given Eq.~(\ref{eq:emp}), in the limit of small-size voxels compared to the simulation cube, we compute $\tilde{\varepsilon}_{\nu,Q~\rm{\rm or}~U}$ at the $i$-th slice as 
\begin{equation}\label{eq:emq}
    \tilde{\varepsilon}_{\nu,Q} (\vec{r}_{i}) = \varepsilon_{\nu,P} (\vec{r}_{i}) \cos{2 \left [\varphi(\vec{r}_{i})+\delta RM_{i} \left(\frac{c}{\nu}\right)^2 \right ]}  
\end{equation}
and 
\begin{equation}\label{eq:emu}
    \tilde{\varepsilon}_{\nu,U} (\vec{r}_{i}) = \varepsilon_{\nu,P} (\vec{r}_{i}) \sin{2 \left [\varphi(\vec{r}_{i})+\delta RM_{i} \left(\frac{c}{\nu}\right)^2 \right ]},  
\end{equation}
where $\varphi$ is the intrinsic polarization angle (perpendicular to $\vec{B}_{\perp}(\vec{r}_{i})$) and $\delta RM_i$ is the specific rotation measure ($RM$) in units of rad m$^{-2}$ defined as
\begin{equation}\label{eq:rms}
    \delta RM_i = 0.81 \int_{\vec{r}_i}^{\vec{r}_{i-1}} \frac{n_e(\vec{r})}{[{\rm cm^{-3}}]} \frac{\vec{B}\cdot {\rm d} \vec{r}}{[\mu {\rm G}][{\rm pc}]}.  
\end{equation}

In our case, the LOS, $\vec{r}$, always represents one of the coordinate axes of the cubes, $\hat{x}$ or $\hat{y}$ or $\hat{z}$. The frequency maps of $Q_{\nu}$ and $U_{\nu}$ result from integrating Eqs.~(\ref{eq:emq}) and (\ref{eq:emu}) along the full length of the simulated cubes. From the Stokes parameters we also derive the polarized intensity $PI_{\nu} = \sqrt{Q_{\nu}^2+U_{\nu}^2}$. 
 
Finally, we convolve the Stokes $I_{\nu}$, $Q_{\nu}$, and $U_{\nu}$ maps with a Gaussian beam of arbitrary full width half maximum (FWHM), simulating the point spread function (PSF) of real observations. This imposes a certain angular resolution on our synthetic data and allows us to include beam depolarization effects. After that, we apply {\tt{rm-synthesis}\footnote{\url{http://github.com/brentjens/rm-synthesis}}}~\citep{Brentjens2005} on $Q_{\nu}$ and $U_{\nu}$ maps to perform Faraday tomography. This allows us to study polarized emission as a function of Faraday depth, $\phi$. $PI_{\phi}$, $Q_{\phi}$ and $U_{\phi}$ are often referred to as Faraday spectra in polarized intensity, Stokes $Q$, or $U$, respectively. 
 

\begin{figure*}[!h]
\begin{center}
\resizebox{0.9\hsize}{!}{\includegraphics{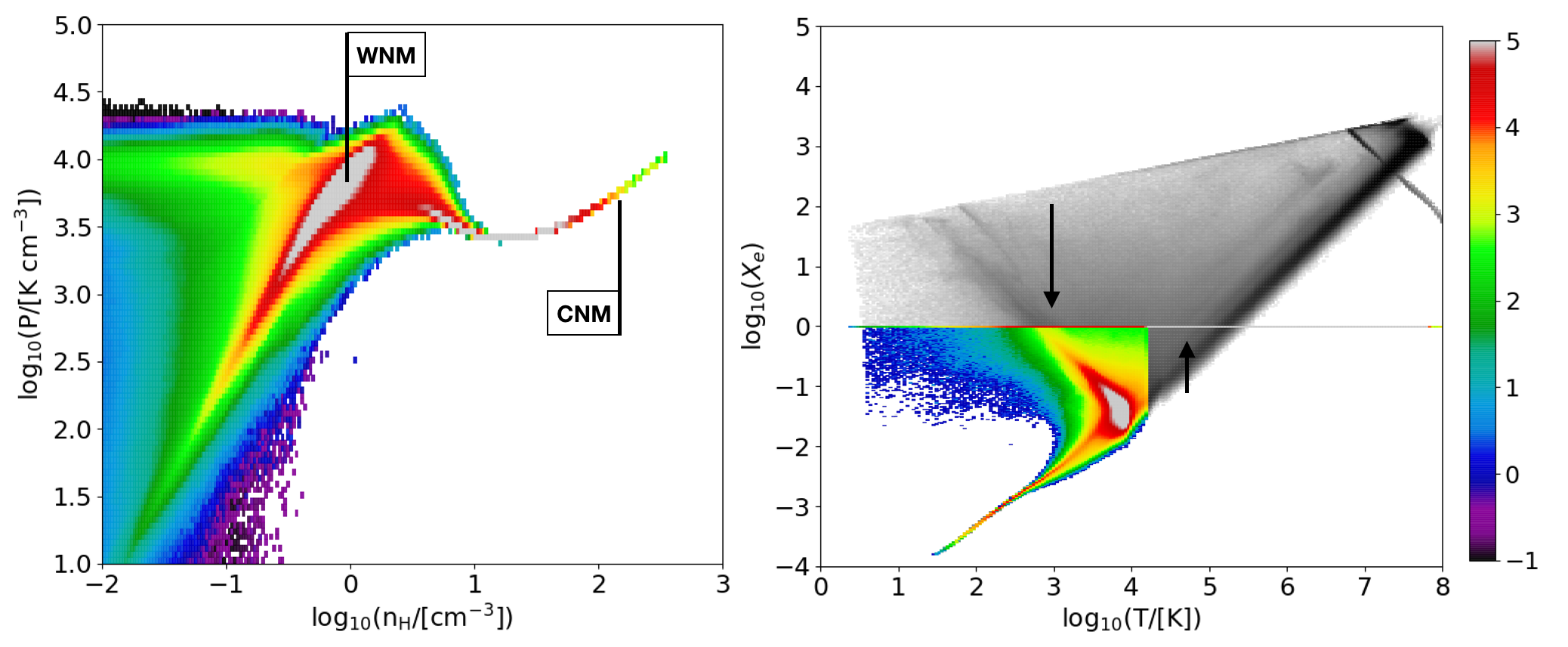}}
\caption{{\it Left panel}: phase diagram (pressure, $P$, vs gas density, $n_{\rm H}$) of case A. The WNM and CNM regions are labeled. {\it Right panel}: dependence of the ionization fraction, $X_{e}$, obtained with Eq.~(\ref{eq:ne}), vs the gas temperature, $T$. Here, the CR ionization rate is set to the value of $\zeta^{\rm H}$ (see Sect.~\ref{ssec:ne}). Colours in both panels correspond to the $\log_{10}$ of the density of points as shown by the same colour bar on the right. The gray part of the right-panel plot shows the voxels that were artificially set to $X_{e} = 1$ (see the two black arrows) as they strongly depart from the assumptions that justify the use of Eq.~(\ref{eq:ne}) (see main text).}
\label{fig:phasediag}
\end{center}
\end{figure*}
 
Since in this work we aim to model synchrotron emission at radio frequencies below 200 MHz,  the synthetic data are tailored to the LOFAR observations \citep[e.g.,][]{Jelic2014,Jelic2015,vanEck2017}. Synchrotron emission in total and polarized intensity is therefore modelled at frequencies from 115 MHz to 170 MHz with steps of 0.18 MHz. This frequency range gives a resolution in Faraday depth of $\sim$1 rad m$^{-2}$, defined by the width of the rotation measure spread function (RMSF, see Fig.\ref{fig:rmsf}). The presented synthetic Faraday cubes span between $-50$ and $+50\,{\rm rad\, m}^{-2}$ in steps of 0.25 rad m$^{-2}$. The maximum observable scale in $\phi$ space is $\pi/\lambda_{\rm min}^2 \sim 1$ rad m$^{-2}$. Any Faraday depth structure along the LOS with an extent in Faraday space larger than $\sim 1$ rad m$^{-2}$ is referred to as Faraday thick in our synthetic data \citep[see][for more details]{Brentjens2005}.  

\subsection{Description of the simulations}\label{ssec:sims}

As a characteristic region of the multiphase ISM, we use MHD simulations of two colliding super-shells (\citealt{Ntormousi2017}, hereafter N17). The super shells are created by placing two spherical feedback regions on either side of a 200 pc box, which is initially filled with a turbulent medium of mean density $n_{\rm H}$=1 cm$^{-3}$ and mean temperature of 8000~K. In order to set up the turbulence, before introducing the feedback, N17 imposed a turbulent velocity field to a box of uniform density and a constant magnetic field along one direction, and let the turbulence evolve until the density-weighted power spectra of the velocity field reach a Kolmogorov-like behavior.
Then the feedback regions were placed on either side of the $z$-axis boundaries, with the magnetic field oriented either perpendicular (case A, \emph{mhd1r} in N17) or parallel (case B, \emph{mhd1t} in N17) to the collision axis. In these regions, the gas receives thermal energy from the combined wind and supernova feedback of an OB association containing 30 stars, following the population synthesis models of \citet{Voss09}.
All the cases used in this work have a uniform resolution of 512$^3$ cells. 
For both cases A and B we use the simulation output at 5 Myrs. This time-step was chosen to have the shells significantly close to each other while still preserving some of the surrounding medium. All gas phases co-exist in the computational box.   
Table \ref{tab:setup} contains the characteristic parameters of the MHD simulations. Self-gravity is not at play in any of the two models (hydrodynamical simulations of the same setup with self-gravity produced very similar results). 

To capture the thermal instability that will eventually create the CNM, N17 also modelled the cooling and heating processes of the local ISM. Heating comes from an UV background modelled with a Habing field of $G_{\rm eff}$=1.7 and from the photoelectric effect on dust grains. Cooling is due to atomic lines, predominantly carbon and oxygen.  The equilibrium rates for cooling and heating were introduced in a tabulated form as a function of density
and temperature for a gas of solar metallicity \citep{Wolfire1995}. 
Further details on the simulations can be found in N17.



\begin{table}[!h]
\caption{Summary of the properties of the $512^3$ simulated cubes from \citet{Ntormousi2017}.}
\begin{center}
\begin{tabular}{ccc}
\toprule\toprule
Case & A & B \\
Initial magnetic field ($\vec{B}_0$) & $(5.0\,\mu {\rm G})\, \hat{y}$ & $(5.0\,\mu {\rm G})\, \hat{z}$ \\
Shell-collision axis & $\hat{z}$ & $\hat{z}$ \\
Time-step [Myrs] & 5 & 6 \\
Initial gas density $n_{\rm H}$ [cm$^{-3}$] & 1 & 1 \\
Initial Temperature [K] & 8000 & 8000 \\

\bottomrule
\end{tabular}
\end{center}
\label{tab:setup}
\end{table}%

\subsection{Estimate of the electron density}\label{ssec:ne}

A key aspect of this work is establishing a proxy of $n_e$ in the multiphase (not isothermal) gas. For gas temperatures $T > 10^{4.2}$ K \citep[][]{Koyama2002,Kim2008,Kim2017}, we consider the gas to be fully collisionally ionized with $n_e = n_{\rm H}$ and an ionization fraction $X_{e} = n_e/n_{\rm H} = 1$. For colder gas, under the assumption of steady-state chemistry for electron abundances in the diffuse ISM, we use the analytical approach introduced by \citet{Wolfire2003} and \citet{Bellomi2020} and deduce $n_e$ from the following parametric formula (see their Eq.~C15 and B.1, respectively):
\begin{equation}\label{eq:ne}
    \frac{n_{e}}{{\rm cm^{-3}}} \approx 2.4\times10^{-3}\left ( \frac{\zeta}{10^{-16}\, {\rm s^{-1}}}\right )^{0.5} \left ( \frac{T}{100\, {\rm K}}\right )^{0.25}\frac{G^{0.5}_{\rm eff}}{\omega_{\rm PAH}} + n_{\rm H}X_{\rm C^+},
\end{equation}
where $\zeta$ is the total ionization rate per hydrogen atom caused by energetic photons (EUV and soft X-ray) and CRs, $\omega_{\rm PAH}$ is the recombination
parameter of electrons onto small dust grains (polycyclic aromatic hydrocarbons, PAH), and $X_{{\rm C}^+}$ is the abundance of ionized carbon, ${\rm C}^+$, relative to $n_{\rm H}$. 
Table~\ref{tab:param} describes the values assigned to all parameters entering Eq.~(\ref{eq:ne}). The limits of our assumptions in the estimate of $n_e$ will be discussed in Sect.~\ref{sec:discussion}.

\begin{table}[!h]
\caption{Parameters for the analytical expression of the electron density $n_e$, see Eq.~(\ref{eq:ne}). Notes: $^{(a)}$input radiation field used in \citet{Ntormousi2017}; $^{(b)}$discussed in \citet{Wolfire2003}; $^{(c)}$value derived in
the Solar Neighborhood assuming 40\% depletion of carbon onto grains \citep{Bellomi2020}.}
\begin{center}
\begin{tabular}{cc}
\toprule\toprule
Parameter  &  Value  \\%
\midrule
$\zeta$ [s$^{-1}$] & $\zeta^{\rm L} = 1.7 \times 10^{-17}\, {\rm or}\,\, \zeta^{\rm H} = 2.6 \times 10^{-16}$\\
$G_{\rm eff}$ [Habing]$^{(a)}$ & 1.7\\
$\omega_{\rm PAH}$$^{(b)}$ & 0.5\\
$X_{{\rm C}^+}$$^{(c)}$& $1.4\times 10^{-4}$\\

\bottomrule
\end{tabular}
\end{center}
\label{tab:param}
\end{table}%

\begin{table*}[!h]
\caption{Criteria in temperature ($T$) and ionization fraction ($X_{e}$) that define the different gas phases. The derived mean gas density ($\bar{n}_{\rm H}$) and its standard deviation ($\sigma_{n_{\rm H}}$) are listed in the fourth column. Indications from \citet{Heiles2012} and \citet{Ferriere2020} were followed.}
\begin{center}
\begin{tabular}{cccc|c}
\toprule\toprule
Gas phase & Acronym &$T$ [K] & $X_{e}$ & $\bar{n}_{\rm H} \pm \sigma_{n_{\rm H}}$  [cm$^{-3}$]\\%
\midrule
 Cold neutral medium & CNM & $<300$ & $<10^{-3}$ & $22\pm 12$\\
 Lukewarm neutral medium & LNM  & $[300,5000)$ & $[10^{-3},10^{-2})$ & $4\pm 2$\\
 Warm neutral medium & WNM & $(10^3,10^4)$ & $[10^{-2},5\times 10^{-2})$& $0.8\pm 0.3$\\
Warm partially ionized medium & WPIM & $(10^3,10^4)$ & $[5\times 10^{-2},1)$ & $0.3 \pm 0.1$\\
 Fully ionized medium & FIM & $> 9000$ & 1 & $0.01 \pm 0.05$ \\
\bottomrule
\end{tabular}
\end{center}
\label{tab:phases}
\end{table*}%

Among these parameters $\zeta$ is the most critical one. Given the average gas column density ($N_{\rm H}$) in the simulations ($\sim 10^{20}$ cm$^{-2}$) the contribution from energetic photons  to $\zeta$ can be as low as two orders of magnitude less than that from CRs \citep[see Table 1 in][]{Wolfire2003}. We thus focus on the contribution of the CR ionization rate. In particular, the CR ionization rate has been observed to vary over more than one order of magnitude in the diffuse ISM \citep[e.g.,][]{Padovani2009,Padovani2018}.
We include two scenarios: (i) we consider a conservative value for the CR ionization rate, such that $\zeta = 1.7 \times 10^{-17}\, {\rm s^{-1}}$ \citep[hereafter $\zeta^{\rm L}$, see][]{Wolfire2003}; (ii) we consider a larger value of $\zeta = 2.6 \times 10^{-16}\, {\rm s^{-1}}$ (hereafter $\zeta^{\rm H}$)\footnote{The superscripts L and H refer to "low" and "high", respectively.}, which better corresponds to recent measurements of the CR ionization rate in the diffuse ISM based on ionized species such as OH$^+$, H$_2$O$^+$, and $\rm H_3^+$ \citep[i.e.,][]{Shaw2008,Neufeld2010,Indriolo2012,Neufeld2017}.

\subsection{Definition of distinct gas phases}\label{ssec:phases}
Depending on the choice of $\zeta$ and $T$, we can distinguish between several components in the simulated multiphase gas. Table~\ref{tab:phases} explicates the criteria used to segment the simulated cubes in distinct gas phases, which differ in terms of temperature and ionization fraction. We use standard nomenclature to refer to most gas phases (CNM, LNM, WNM, WPIM) except for what we call fully ionized medium (FIM), which includes both warm (WIM) and hot (HIM) ionized gas in the simulations (see Sect.\ref{sec:intro}).  
\begin{figure}[!h]
\begin{center}
\resizebox{0.9\hsize}{!}{\includegraphics{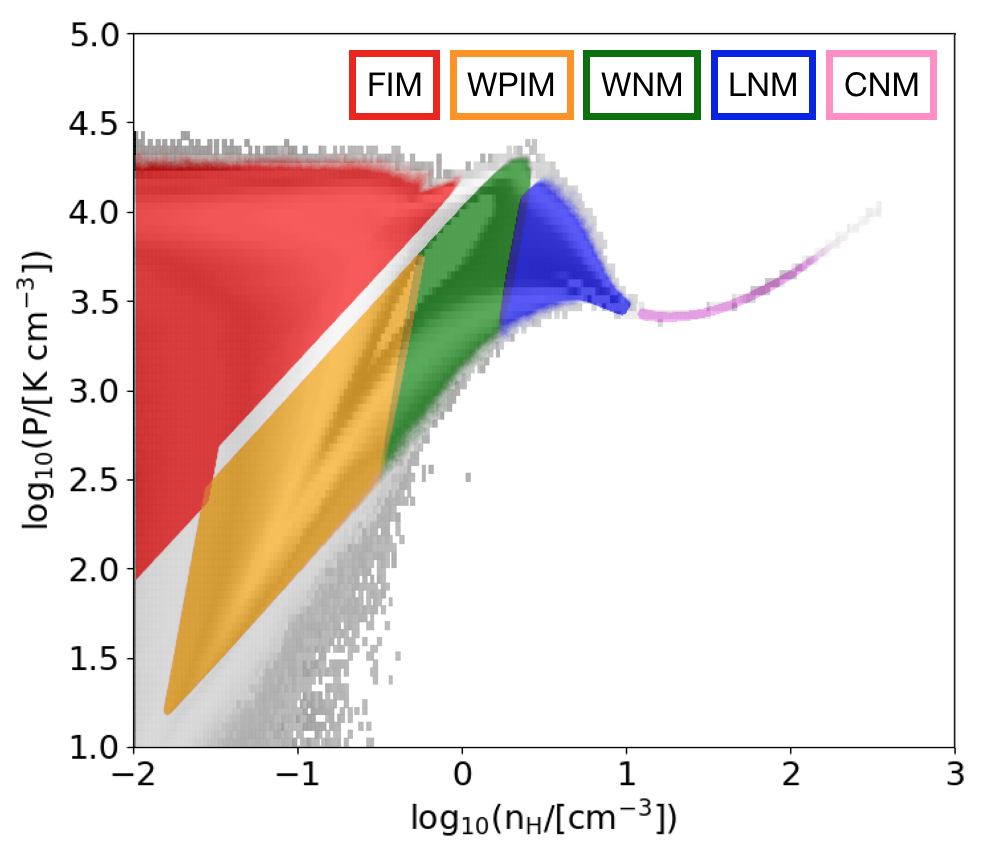}}
\caption{Same as in the left panel of Fig.~\ref{fig:phasediag} for case A but with colours delimiting the regions corresponding to each gas phase as defined in Table~\ref{tab:phases}.}
\label{fig:phasediag2}
\end{center}
\end{figure}
\begin{figure*}[!h]
\begin{center}
\vspace{1.5cm}
\resizebox{\hsize}{!}{\includegraphics{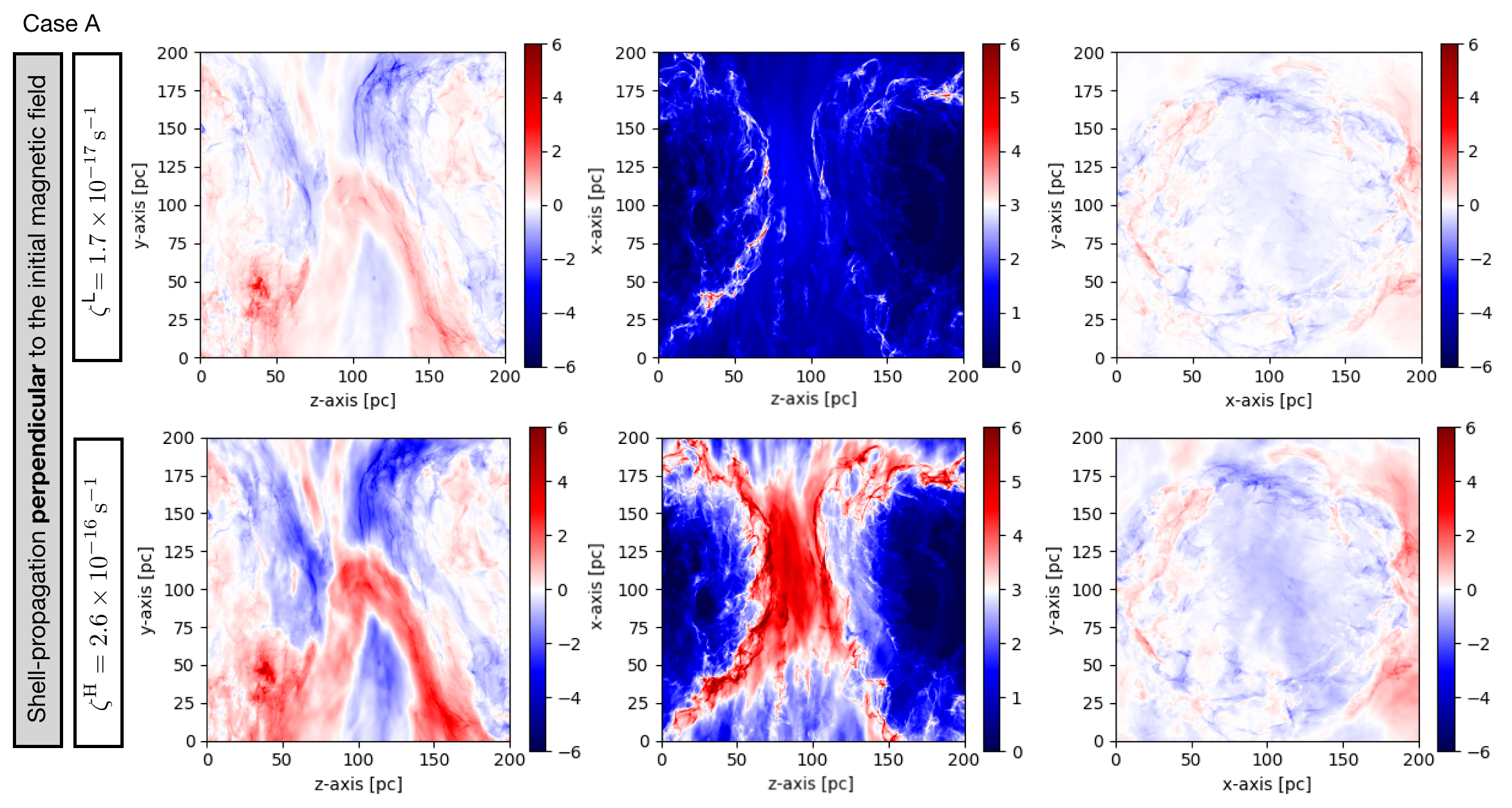}}
\resizebox{\hsize}{!}{\includegraphics{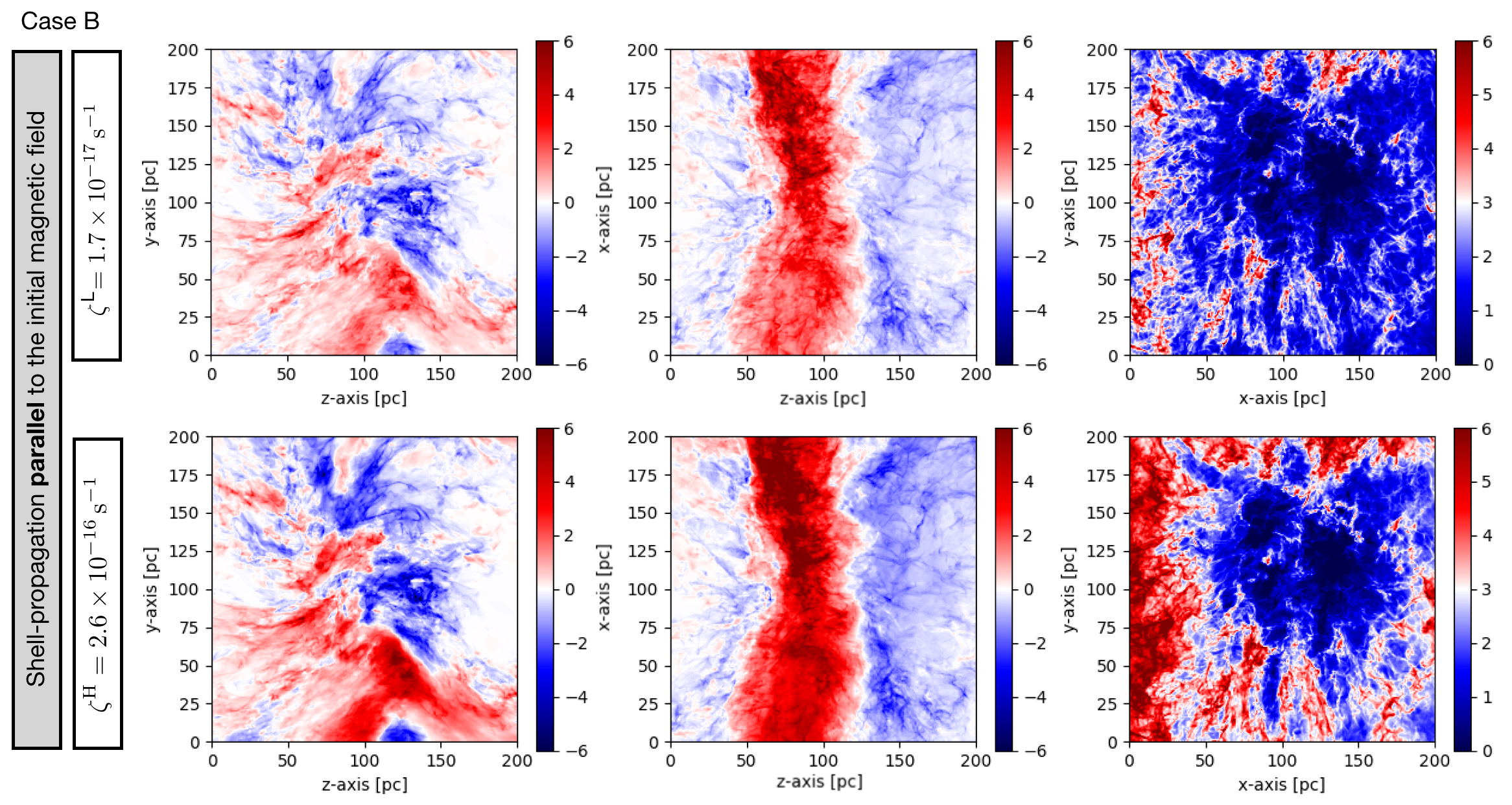}}
\caption{Maps of rotation measure ($RM$) in units of rad m$^{-2}$ obtained within the full length of the simulated cube for cases A and B with two amounts of ionization rates ($\zeta$, see encapsulated panels on the left). The LOS changes along the $x$ or $y$ or $z$ axes going from left to right, respectively. The dynamic range of the color bars is only positive when the integration axis is along the mean magnetic-field orientation (central panels in the first two rows from the top and right panels in the two bottom rows).}
\label{fig:rmtot}
\end{center}
\end{figure*}
The left panel of Fig.~\ref{fig:phasediag} shows the phase-diagram of pressure, $P$, against $n_{\rm H}$ for case A. The typical branches of WNM and CNM (regions where $P$ increases isothermally with $n_{\rm H}$) can be seen; as well as the unstable LNM phase in between (see also Fig.~\ref{fig:phasediag2}). For these phases Eq.~(\ref{eq:ne}) applies. The simulations also contain a large fraction of gas that is more diffuse and ionized than the standard WNM. In the right panel of Fig.~\ref{fig:phasediag} we show how, given the value of $\zeta^{\rm H}$, $X_e$ changes with $T$. The grey-scale shows the voxels in the simulation that we force to be fully ionized to circumvent values of $X_e > 1$. This highlights the limits of Eq.~(\ref{eq:ne}) to analytically infer $n_e$ in our models.
Nevertheless, for the largest fraction of voxels (in colours) $X_e < 1$. 

The relation between $X_e$ and $T$ is not a trivial and monotonic function. The spread of $X_e$ is such that gas at typical WNM temperatures can be highly ionized in the simulation. According to our definition, this phase corresponds to the WPIM. Figure~\ref{fig:phasediag2} displays in colours the regions delimiting all selected phases overlaid on the phase-diagram shown in Fig.~\ref{fig:phasediag}, with corresponding mean gas densities ($\bar{n}_{\rm H}$) listed in the right column of Table~\ref{tab:phases}. In case A, 97\% percent of the voxels have conditions corresponding to at least one of the phases in Table~\ref{tab:phases}; in case B, 96\%. The volume fraction of each phase corresponding to the two cases is listed in Table~\ref{tab:volfrac}.

\begin{table}[!h]
\caption{Volume fractions per gas phase depending on the model.}
\begin{center}
\begin{tabular}{cccccc}
\toprule\toprule
Case & CNM+LNM & WNM & WPIM & FIM & Total \\%
\midrule
 A & 1.5\% & 33.5\% & 23\% & 39\%& 97\% \\
 B & 6\% & 14\% & 8\% & 68\%& 96\%  \\
\bottomrule
\end{tabular}
\end{center}
\label{tab:volfrac}
\end{table}%

\section{Results}\label{sec:results}

In this section we present the main results of our work based on the analysis of synthetic observations of Faraday rotation and tomography. In Sect.~\ref{ssec:rmmaps} we present maps of $RM$ depending on the choice of $\zeta$. Section~\ref{ssec:rmneb} shows the link between the maps of $RM$ and the structures of electrons and magnetic fields in the simulations. In Sect.~\ref{ssec:phaserm} we investigate the contribution of each gas phase (as defined in Table~\ref{tab:phases}) to the map of $RM$. In Sect.~\ref{ssec:tomo} we present the mock observations of Faraday tomography, while in Sect.~\ref{ssec:phasepi} we explore the contribution of each gas phase to the amount of detectable synchrotron polarized intensity below 200 MHz. 
 
\subsection{Maps of rotation measure}\label{ssec:rmmaps}
The choice of $\zeta$ plays a key role in determining the amount of ionized gas in the simulations. This is nicely seen in the maps of total $RM$ that we show in Fig.~\ref{fig:rmtot}. The figure displays $RM$ computed for cases A and B, which shows the Faraday depth integrated across the full 200
pc length of the cubes. 
The integration along $x$, $y$, and $z$ is shown from left to right. The super-shells can be seen colliding edge-on in the former two cases, while face-on in the latter.
Regardless of the integration axis, we obtain a wider range of $RM$ values using $\zeta^{\rm H}$ compared to $\zeta^{\rm L}$ because of the overall larger amount of ionized gas.

\begin{figure}[!h]
\begin{center}
\resizebox{1\hsize}{!}{\includegraphics{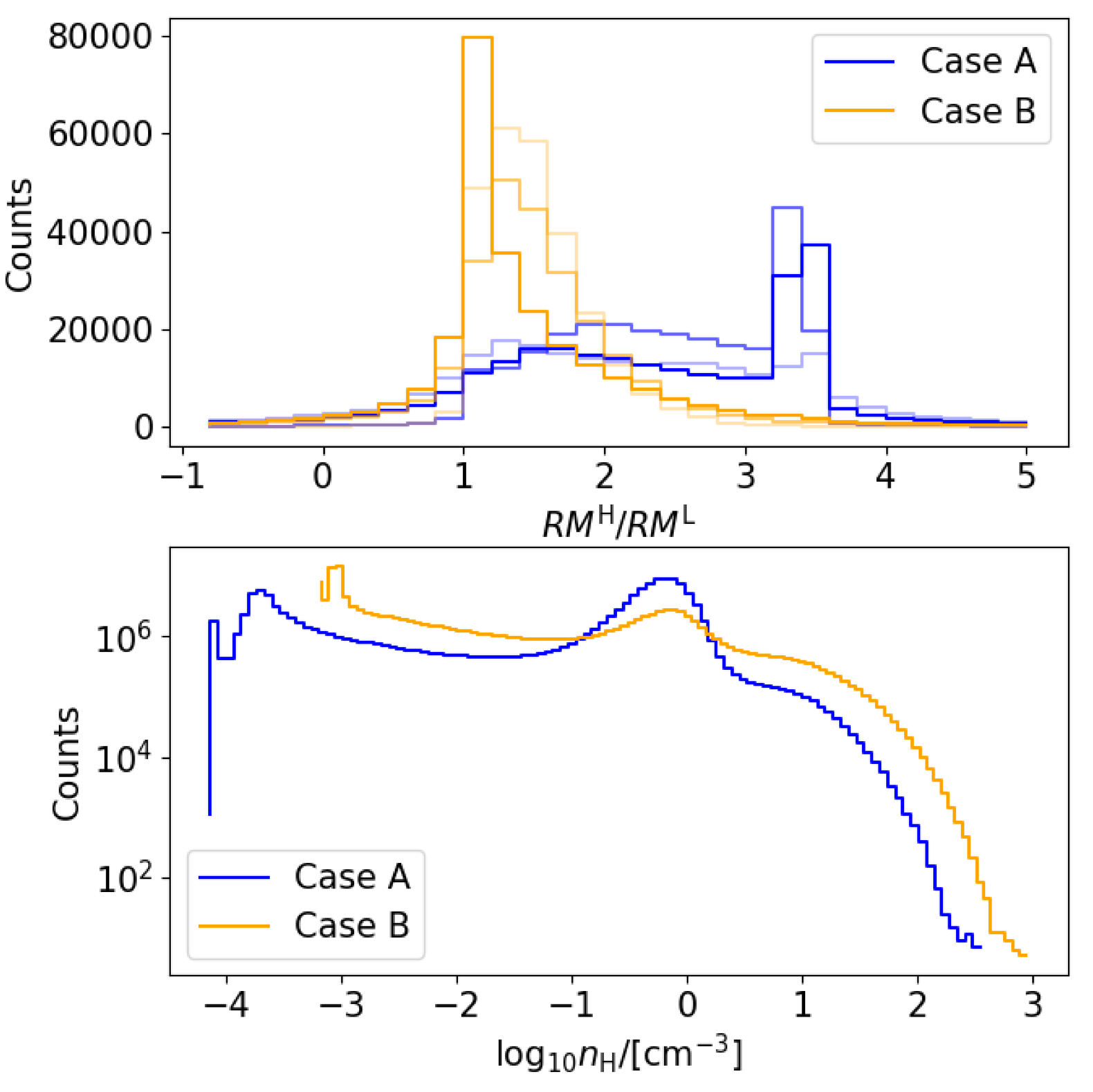}}
\caption{Top: Histograms of the ratios of $RM$ obtained with a high ionization rate ($\zeta^{\rm H}=2.6\times10^{-16}$~s$^{-1}$) and low  ionization rate ($\zeta^{\rm L} = 1.7\times 10^{-17}$ s$^{-1}$) for cases A and B, in blue and orange, respectively. Histograms with different transparencies correspond to integration axes, $x$, $y$, and $z$, from thick to light curves, respectively. Bottom: histograms of the gas density for cases A and B.}
\label{fig:ionratios}
\end{center}
\end{figure}

The choice of the integration axis has a strong impact on the distribution of $RM$ values. In all cases the structure in the maps appears as a mixture of large-scale and small-scale filamentary structures. The $RM$ range covers both negative and positive values
when the LOSs are perpendicular to the mean magnetic-field direction (see Table~\ref{tab:setup}), while they have mostly positive values when the LOS is parallel to it. In the former case the $RM$ structure is therefore dominated by the non-regular component of the magnetic field. 

The largest difference between $RM$ maps computed using $\zeta^{\rm H}$ and $\zeta^{\rm L}$, hereafter labeled as $RM^{\rm H}$ and $RM^{\rm L}$, is for case A.
In the top panel of Fig.~\ref{fig:ionratios} we show the histograms of the ratios between $RM^{\rm H}$ and $RM^{\rm L}$ for all integration axes. It is clear that, if on the one hand, in case B, $RM^{\rm H}/RM^{\rm L} \approx 1$, on the other hand case A shows $RM^{\rm H}/RM^{\rm L} > 1$ with a peak between 3 and 4. 

The difference between cases A and B can be explained in terms of the amount of dense gas (see Table~\ref{tab:volfrac}). The histograms of $n_{\rm H}$ at the bottom panel of Fig.~\ref{fig:ionratios} demonstrate that case B has generally denser media than case A, despite their similar evolutionary time-step. This is mostly because the compression of the gas produced by the super-shell collision in case A is opposed by the magnetic-field tension that acts perpendicular to the collision axis. Case A shows a more prominent peak of WNM (at $n_{\rm H}\approx 1$ cm$^{-3}$) compared to case B, where WNM already turned into CNM at larger density. The choice of $\zeta$ can become crucial depending on the physical configuration of the model, or on the amount of diffuse gas present in the simulation. The more diffuse gas in the simulation, the larger the impact of $\zeta$.
Bearing this in mind, hereafter we use the value of $\zeta^{\rm H}$ (see Sect.~\ref{ssec:ne}), because the range of $RM^{\rm H}$ (roughly between $-10$ rad m$^{-2}$ and $+10$ rad m$^{-2}$) highly resembles the $\phi$-range at which $PI$ is observed with LOFAR in the diffuse ISM within a few hundred parsecs from the Sun \citep[i.e.,][]{Jelic2015,vanEck2017,Bracco2020b,Turic2021}. 

\subsection{Impact of $\vec{B}$ and $n_e$ on rotation measure}\label{ssec:rmneb}

The physical interpretation of $RM$ values, as those shown in the maps above, is complicated by the degeneracy between $n_{e}$ and the LOS-component of $\vec{B}$ (hereafter, $B_{\parallel}$), as well as the path-length along the LOS (see Eq.~(\ref{eq:rms})). While veritably a problematic issue with radio observations of diffuse polarized emission \citep[e.g.,][]{Jelic2014, Lenc2016, vanEck2017, Thomson2019, Turic2021}, the degeneracy between $n_{e}$ and $B_{\parallel}$ can be sometimes circumvented in the case of pulsar measurements \citep[e.g.,][]{Smith1968,Rand1989,Han1999,Han2006,Sobey2019}. In particular, pulsars give access to the dispersion measure ($DM$), defined in units of pc cm$^{-3}$ as $DM = \int_0^d n_e {\rm d}r$, where $d$ is the distance to the pulsar and $r$ is the LOS. Combining $DM$ with $RM$, in units of rad m$^{-2}$, allows one to estimate the density-weighted average strength of $B_{\parallel}$ in units of $\mu$G as follows
\begin{equation}\label{eq:avbpar}
    \langle B_{\parallel} \rangle_{\rm pul} = 1.232 \frac{RM}{DM}.    
\end{equation}

\begin{figure}[!h]
\begin{center}
\resizebox{0.95\hsize}{!}{\includegraphics{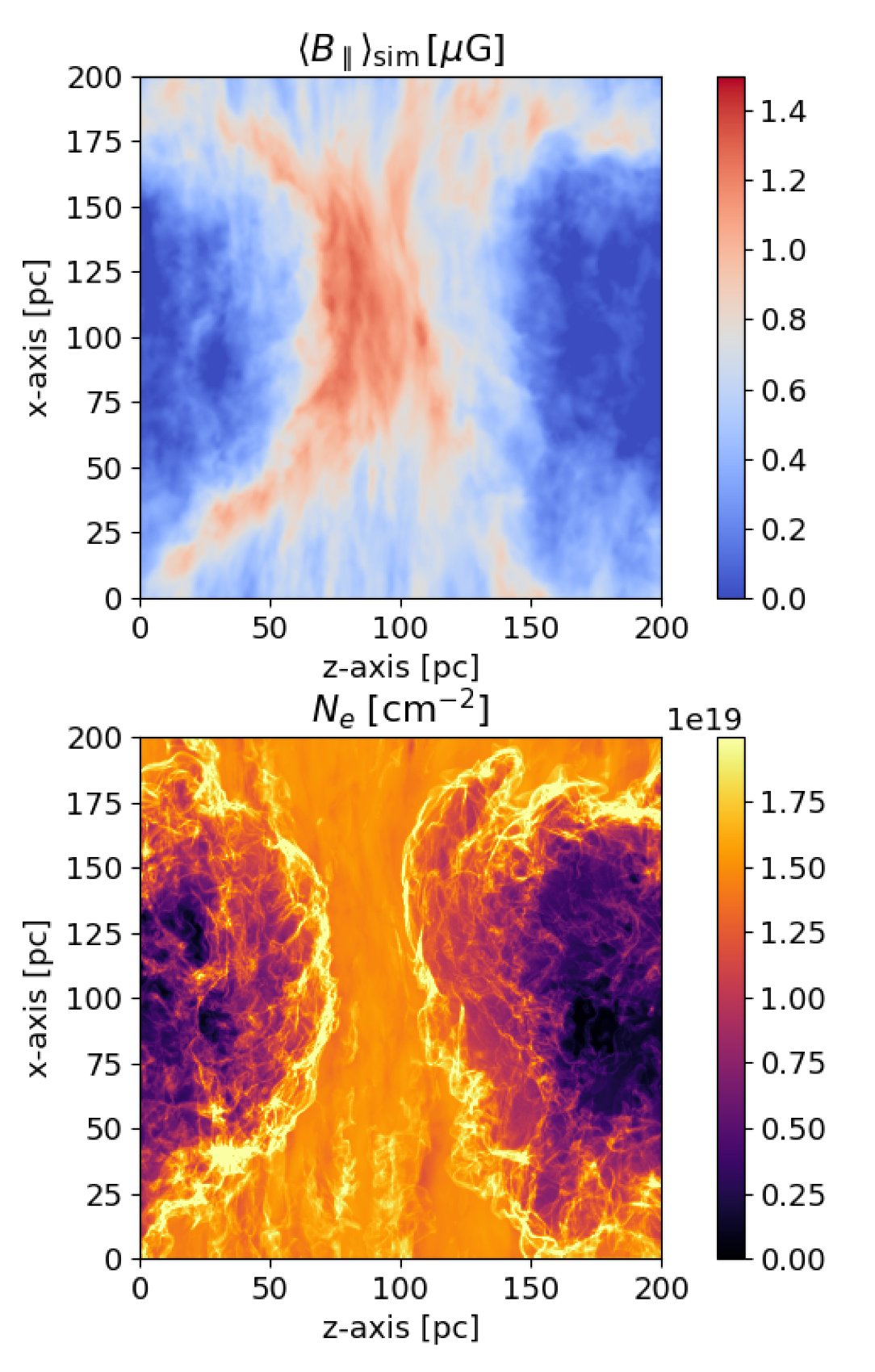}}
\caption{Maps of the LOS-average magnetic field (top) and the electron column density ($N_{e}$, bottom) for case A integrated along the $y$ axis.}
\label{fig:rmbne}
\end{center}
\end{figure}

In this section we investigate our simulations and ask whether we are able to discriminate magnetic fields from electrons in the synthetic observations of $RM^{\rm H}$. 
In Fig.~\ref{fig:rmbne} we show maps of the LOS-average of $B_{\parallel}$ (hereafter, $\langle B_{\parallel}\rangle_{\rm sim}$) computed along the $y$ axis in case A, as well as the corresponding electron-column density ($N_{e}$).  

\begin{figure}[!h]
\begin{center}
\resizebox{\hsize}{!}{\includegraphics{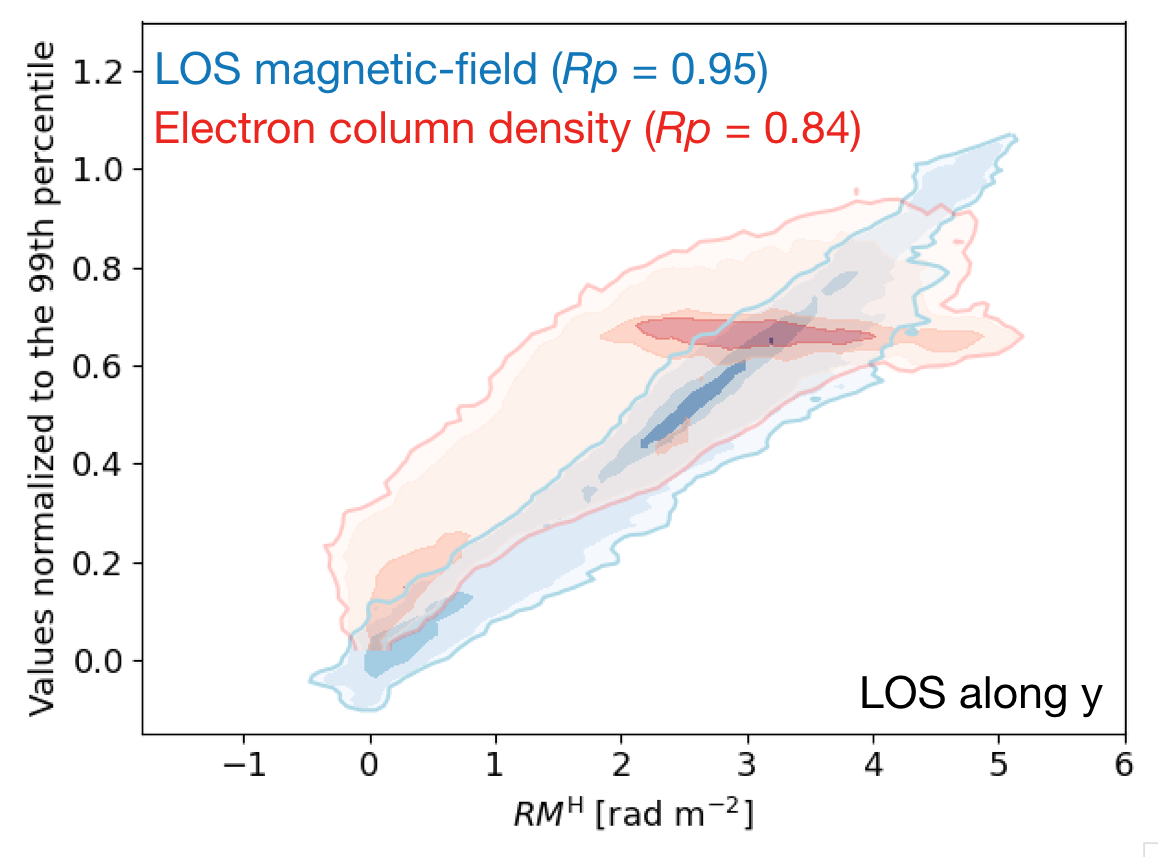}}
\caption{2D histograms showing the correlation between $RM^{\rm H}$ (see Fig.\ref{fig:rmtot}), and the maps of the LOS-average magnetic field (blue) and the electron column density (red). The ordinate axis shows the corresponding values normalized to their 99th percentile. From light to dark colours, contours correspond to number of pixels of 50, 100, 500, 1000, 2000. Person correlation coefficients ($R_{\rm p}$) are written. As an example, we show case A integrated along the $y$ axis.}
\label{fig:rmbnecor}
\end{center}
\end{figure}

Visually, the map of $RM^{\rm H}$ (see central panel in the second row from the top of Fig.~\ref{fig:rmtot}) is strongly correlated with the structure of $\langle B_{\parallel}\rangle_{\rm sim}$ and with that of $N_{e}$ mostly toward the densest regions. The Pearson correlation coefficients ($R_{\rm p}$) between $RM^{\rm H}$ and $\langle B_{\parallel}\rangle_{\rm sim}$, or $N_{e}$, are 0.95 and 0.84, respectively. 2D histograms encoding these correlations are shown in Fig.~\ref{fig:rmbnecor}, where the distributions of $\langle B_{\parallel}\rangle_{\rm sim}$ (in blue) and $N_{e}$ (in red) are normalized to their 99th percentile.

In Fig.~\ref{fig:rmbna1} we present the same 2D histograms but for different integration axes and for case B. In all explored scenarios the values of $RM^{\rm H}$ appear tightly correlated with those of $\langle B_{\parallel}\rangle_{\rm sim}$. In the case of $N_{e}$, the correlation measured by $R_{\rm p}$ is generally weaker or absent (see left panels of Fig.~\ref{fig:rmbna1}). It is not negligible when integration axis is along the main magnetic-field orientation.

\begin{figure}[!h]
\begin{center}
\resizebox{\hsize}{!}{\includegraphics{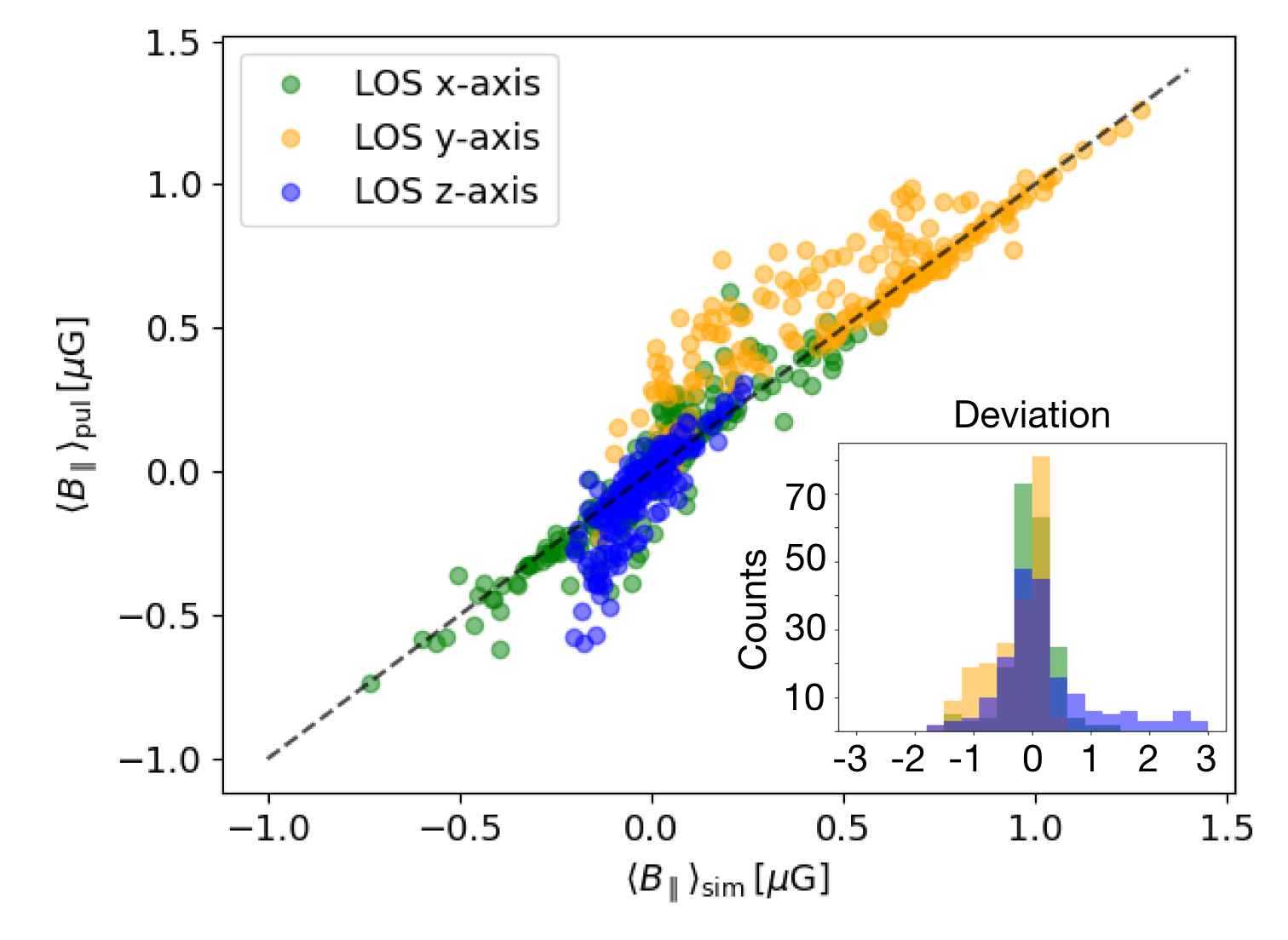}}
\caption{Correlation between $\langle B_{\parallel}\rangle_{\rm sim}$ and $\langle B_{\parallel}\rangle_{\rm pul}$ for case A. A one-to-one dashed line is overlaid. The inset shows the deviation of $\langle B_{\parallel}\rangle_{\rm pul}$ from $\langle B_{\parallel}\rangle_{\rm sim}$ defined as ($\langle B_{\parallel}\rangle_{\rm sim} - \langle B_{\parallel}\rangle_{\rm pul})/\sigma_{\rm sim}$, where $\sigma_{\rm sim}$ is the standard deviation of $\langle B_{\parallel}\rangle_{\rm sim}$.}
\label{fig:puls}
\end{center}
\end{figure}

Finally, we notice that, as shown in Fig.~\ref{fig:puls}, from  Eq.~(\ref{eq:avbpar}) we are able to give a reliable estimate of $\langle B_{\parallel}\rangle_{\rm sim}$ using $\langle B_{\parallel}\rangle_{\rm pul}$, regardless of the integration axis. We produced this plot by accounting for 200 LOSs randomly chosen within the simulated boxes. The number of LOSs is not key to validate Eq.~(\ref{eq:avbpar}). The scatter of the 200 LOSs is comparable along all integration axes and shows that $\langle B_{\parallel}\rangle_{\rm pul}$ and $\langle B_{\parallel}\rangle_{\rm sim}$ are consistent within 1 $\sigma$ (see inset in Fig.~\ref{fig:puls}). 

\subsection{Multiphase gas contribution to rotation measure}\label{ssec:phaserm}
The range of $RM$ depends on the phase distribution of the gas in the simulations (see Sect.~\ref{ssec:phases}). The five phases that we defined above are unevenly distributed in the cubes and show distinct morphology depending on their mean gas density and temperature (see Table~\ref{tab:phases}). 

\begin{figure}[!h]
\begin{center}
\resizebox{0.9\hsize}{!}{\includegraphics{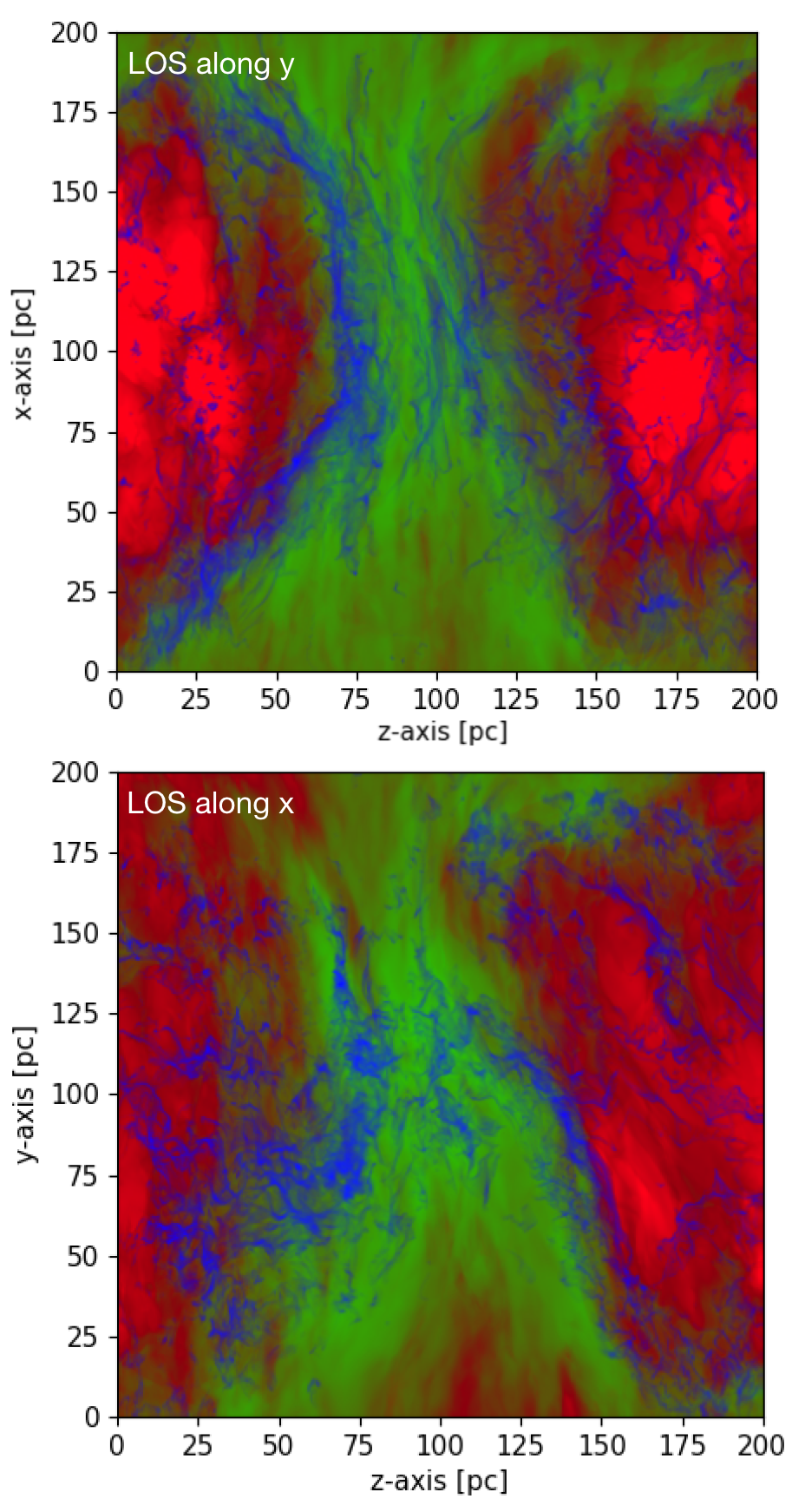}}
\caption{RGB images showing the contribution to the column density structure of the cold phases (blue, CNM+LNM), the warm and partially ionized phases (green, WNM+WPIM), and the fully ionized phase (red, FIM) as defined in Table~\ref{tab:phases}. Two different LOS for case A are shown.}
\label{fig:rgb}
\end{center}
\end{figure}

In Fig.~\ref{fig:rgb} we show RGB images of $N_{\rm H}$ maps corresponding to the cold (CNM+LNM), warm (WNM+WPIM), and hot (FIM) phases for two different integration axes of case A. All phases appear to be affected by the shell collision. While the hot and warm phases are mostly structured on large scale, as expected, the coldest and densest phases show small-scale structures that are the consequence of thermal instability in the WNM. 

The morphology of gas phases is highly correlated and complementary, although not necessarily co-spatial. The edges of each $N_{\rm H}$ map appear aligned among phases. In order to quantify this, we applied histograms of oriented gradients \citep[HOG\footnote{\url{http://github.com/solerjuan/astrohog}},][]{Soler2019} to the $N_{\rm H}$ maps obtained for each distinct phase. The basic principle of HOG is to provide a statistical estimate of the spatial correlation (morphological alignment) between two maps assuming that the local appearance and shape of a map can be fully characterized by the distribution of its local intensity gradients or edge directions. To evaluate the correlation we used the HOG output parameter defined as the projected Rayleigh statistics ($V$, see Eq. C.1 in \citealp{Bracco2020b}). The $V$ parameter is a number that represents the likelihood that the gradients of two maps are mostly parallel. Larger values of $V$ correspond to stronger alignment. As noticed by \citet{Soler2019}, it is not possible to draw conclusions from the values of $V$ alone, but its statistical significance can be assessed by comparing a given $V$ value to others obtained in maps with similar statistical properties. 

Because of the different volume filling fractions of the hot/warm and cold phases (see Table~\ref{tab:volfrac}), the surface area covered by the $N_{\rm H}$ values of the former is significantly larger than the latter. Thus, when computing HOG we normalized our $V$ values to the amount of pixels where the $N_{\rm H}$ of CNM is not zero.   
\begin{figure}[!h]
\begin{center}
\resizebox{\hsize}{!}{\includegraphics{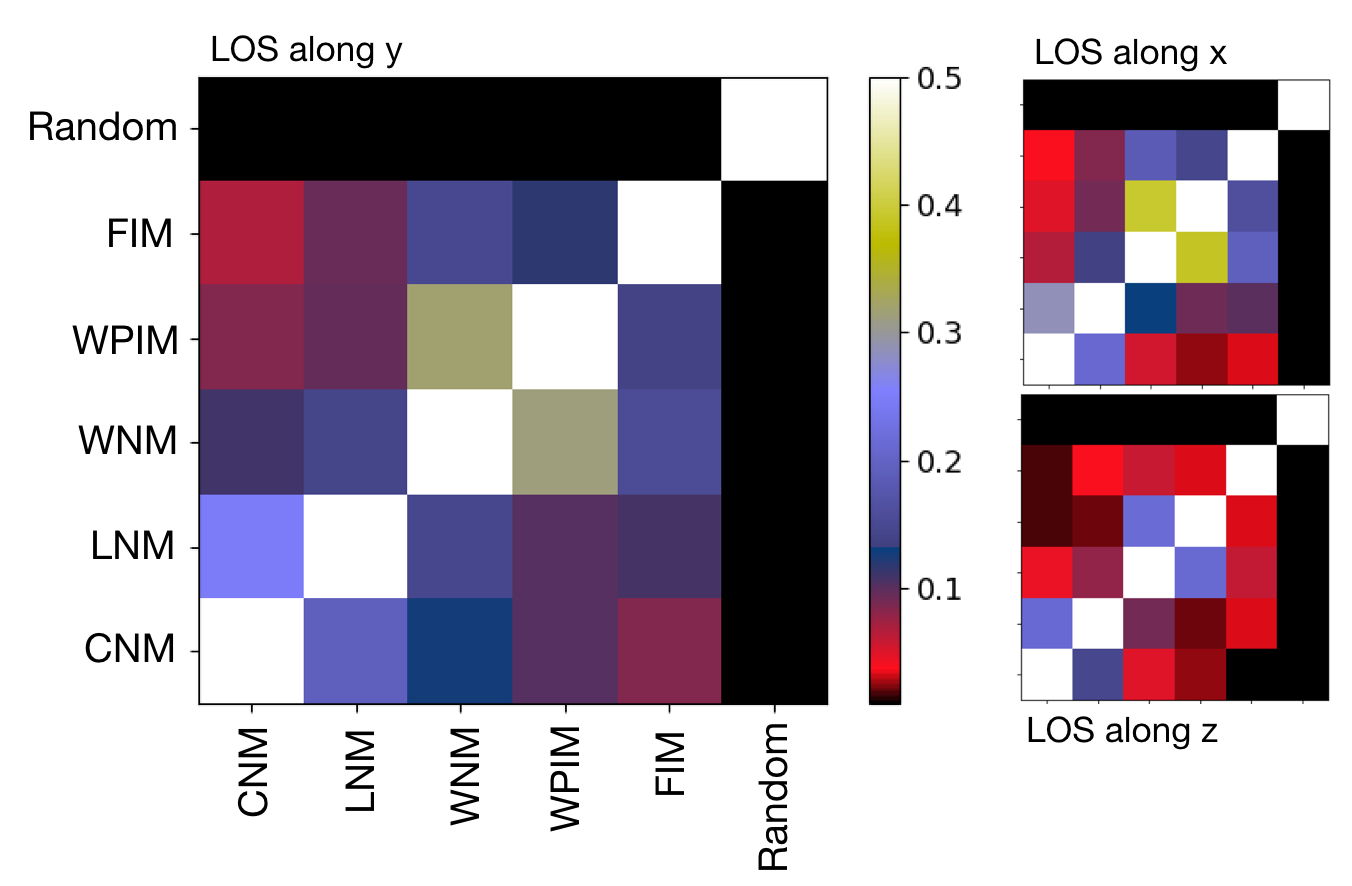}}
\caption{Maps of the normalized projected Rayleigh statistics obtained with HOG (in colours) between the column density map of each phase as labeled in the figure. One random map for reference is also considered. Three different lines of sight of case A are shown. Larger values correspond to a higher degree of correlation.}
\label{fig:hogphase}
\end{center}
\end{figure}
The normalized $V$ values for case A are shown in Fig.~\ref{fig:hogphase}. We studied the correlation among all phases for the three integration axes. 
$V$ correctly shows no correlation with a Gaussian random field (labeled as "Random" in the figure). As shown in Fig.~\ref{fig:hogphase}, the closer phases are in phase space, the more their maps look alike.

\begin{figure}[!h]
\begin{center}
\resizebox{\hsize}{!}{\includegraphics{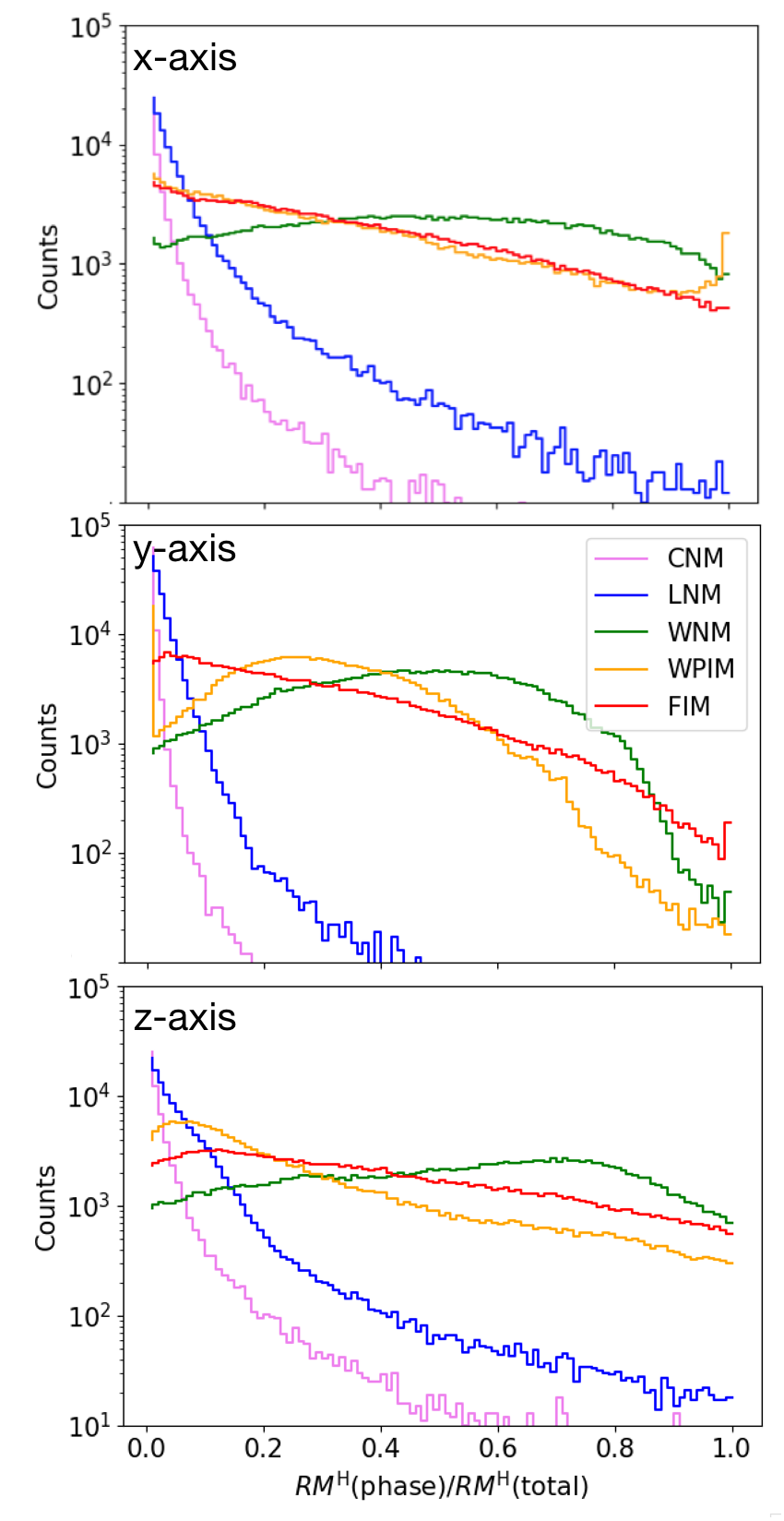}}
\caption{Histograms of the relative contribution to the total rotation measure ($RM^{\rm H}({\rm total})$) of each gas phase ($RM^{\rm H}({\rm phase})$) as defined in Table~\ref{tab:phases} for case A. Colors are defined in the central panel.}
\label{fig:rmrel1}
\end{center}
\end{figure}

This multiphase and multiscale structure in the simulations has an impact on the values of $RM^{\rm H}$, since not all phases contribute the same to the rotation measure. For each gas phase, we computed $RM$ maps using only voxels belonging  to that phase. In Fig.~\ref{fig:rmrel1}, for case A, we show the distributions of the relative contribution of each phase to the total $RM^{\rm H}$. We notice that the share of $RM^{\rm H}$ among phases depends on the integration axis (see also Fig.~\ref{fig:rmrel2} for case B). Moreover, in case A the phases that contribute the most to the total $RM^{\rm H}$ are WNM and WPIM, in spite of their lower $X_{e}$ compared to FIM. The same is not true for case B, where FIM dominates the total $RM^{\rm H}$. In our simulations CNM and LNM are generally found, as expected, to be negligible phases to the rotation measure.      

\subsection{Mock observations of Faraday tomography}\label{ssec:tomo}

\begin{figure}[!h]
\begin{center}
\resizebox{\hsize}{!}{\includegraphics{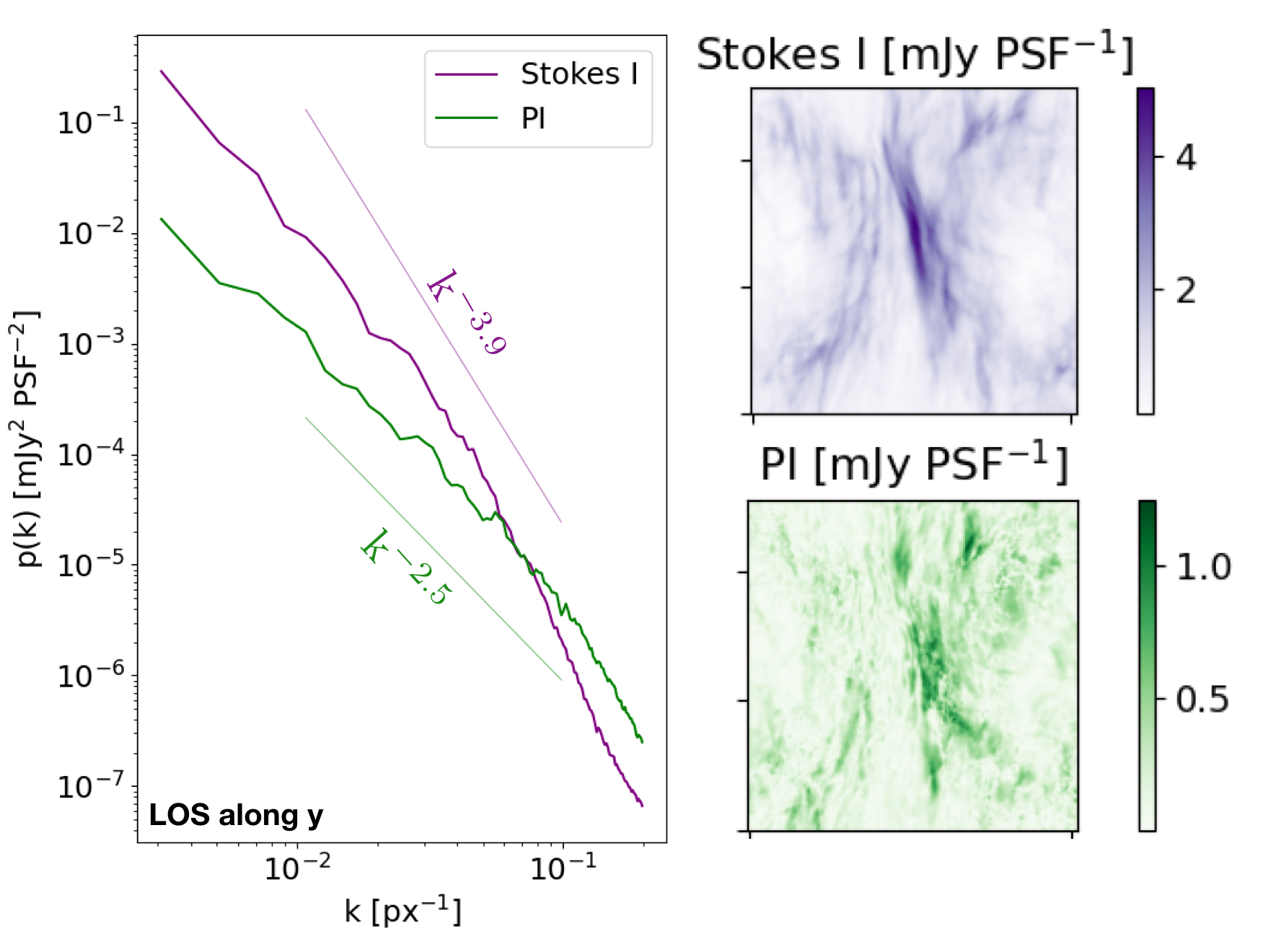}}
\caption{Synthetic observations of synchrotron emission at 150 MHz: maps of Stokes $I$ (top-right panel) and polarized intensity ($PI$, bottom-right panel) with corresponding angular power spectra (left panel). As an example, only case A integrated along the $y$ axis is shown.}
\label{fig:spectra}
\end{center}
\end{figure}

The imprint of each gas phase on the rotation measure also has an impact on the observed synchrotron polarized emission. 
As detailed in Sect.~\ref{ssec:RM}, we produced mock observations of synchrotron emission -- total and polarized --  both as a function of $\nu$ and, through Faraday tomography, of $\phi$. We chose a FWHM of the PSF of a few arcminutes ($\sim$7$\arcmin$) to be roughly comparable with that of LOFAR\footnote{LOFAR observations have a FWHM of $4 \arcmin$ \citep[e.g., ][]{Jelic2014}}, placing the simulated cubes at a hypothetical distance between 500 and 600 pc\footnote{We notice that if this distance was lower, the non-orthonormal projection of the cube should be taken into account. In our case at a distance of 575~pc, the angular size of the voxels changes at most between $2.4\arcmin$ and $1.8\arcmin$.}. In the right panels of Fig.~\ref{fig:spectra} we show, as an example, both Stokes $I$ (top) and $PI$ (bottom) at 150 MHz obtained for case A integrated along the $y$ axis in units of mJy PSF$^{-1}$. At 150 MHz we retrieve, at most, only 20\% of Stokes $I$ in polarization ($PI_{150}/I_{150}$ has a median value of 16\%). The observed depolarization is due to the combined effects of the beam and of differential Faraday rotation in the cubes. The latter also introduces small-scale structure in polarization that is not observed in Stokes $I$. 

\begin{figure*}[!h]
\begin{center}
\resizebox{0.65\vsize}{!}{\includegraphics{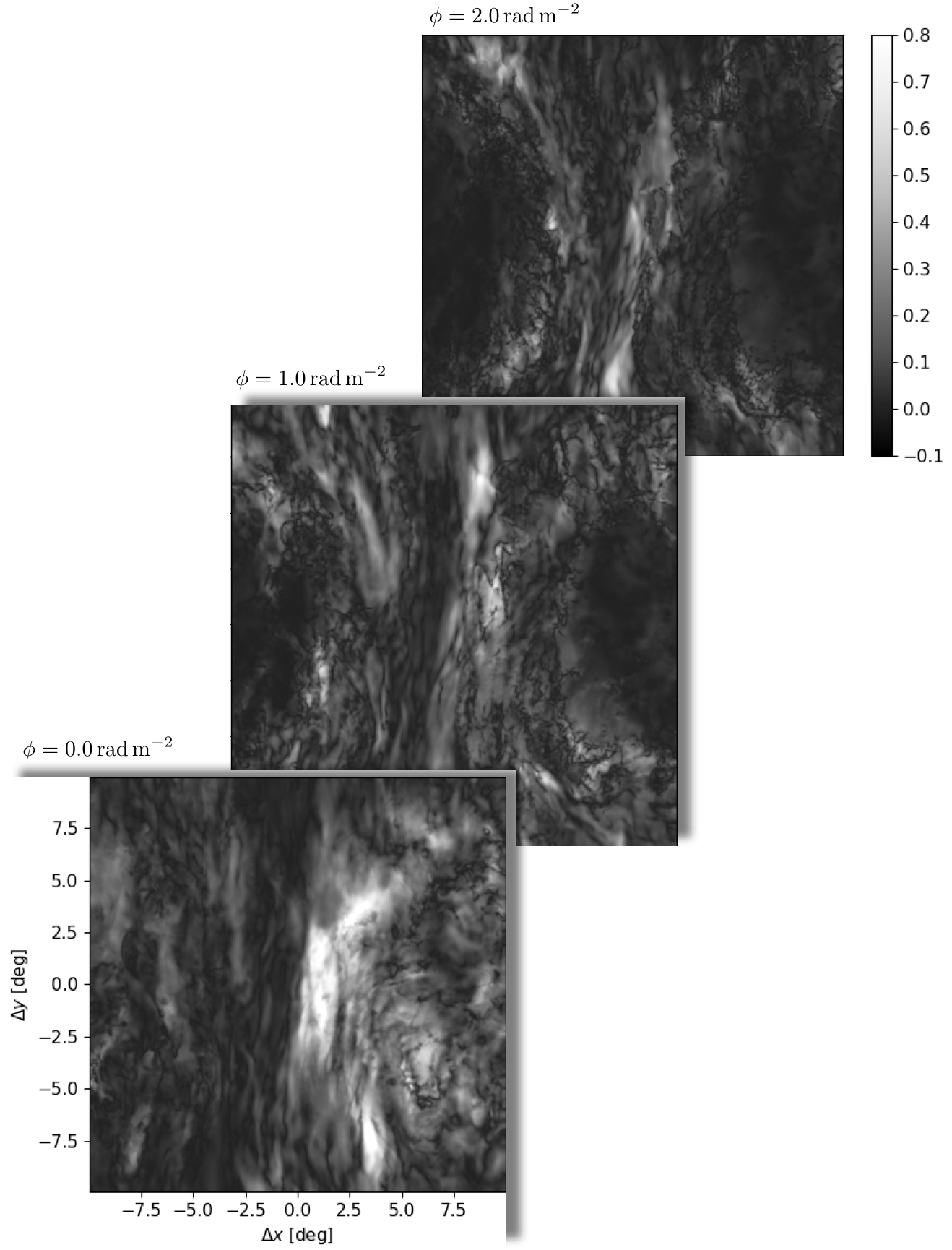}}
\caption{Mock observations of Faraday tomography: maps of $PI$ in units of mJy PSF$^{-1}$ RMSF$^{-1}$ as a function of Faraday depth, $\phi$. The grey scale is the same in all three maps. As an example case A integrated along the $y$ axis is shown.}
\label{fig:tomo}
\end{center}
\end{figure*}

\begin{figure*}[!h]
\begin{center}
\resizebox{\hsize}{!}{\includegraphics{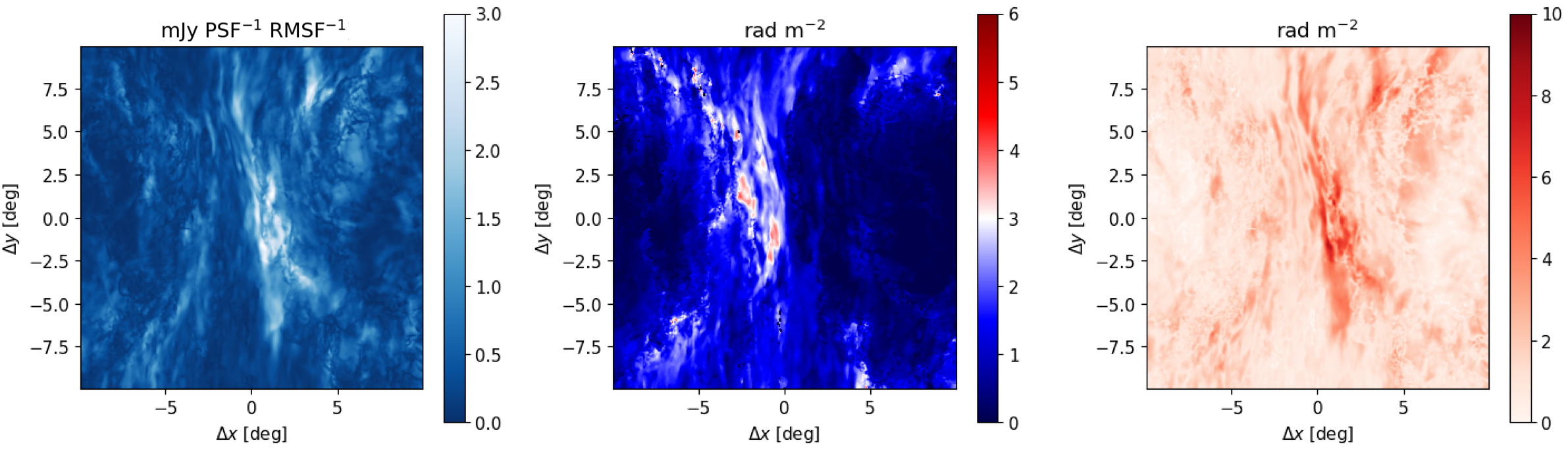}}
\caption{Faraday moments ($M_0$, $M_1$, $\sqrt{M_2}$, from left to right, respectively) of the tomographic cubes shown in Fig.~\ref{fig:tomo}.}
\label{fig:moments}
\end{center}
\end{figure*}

\begin{figure*}[!h]
\begin{center}
\resizebox{\hsize}{!}{\includegraphics{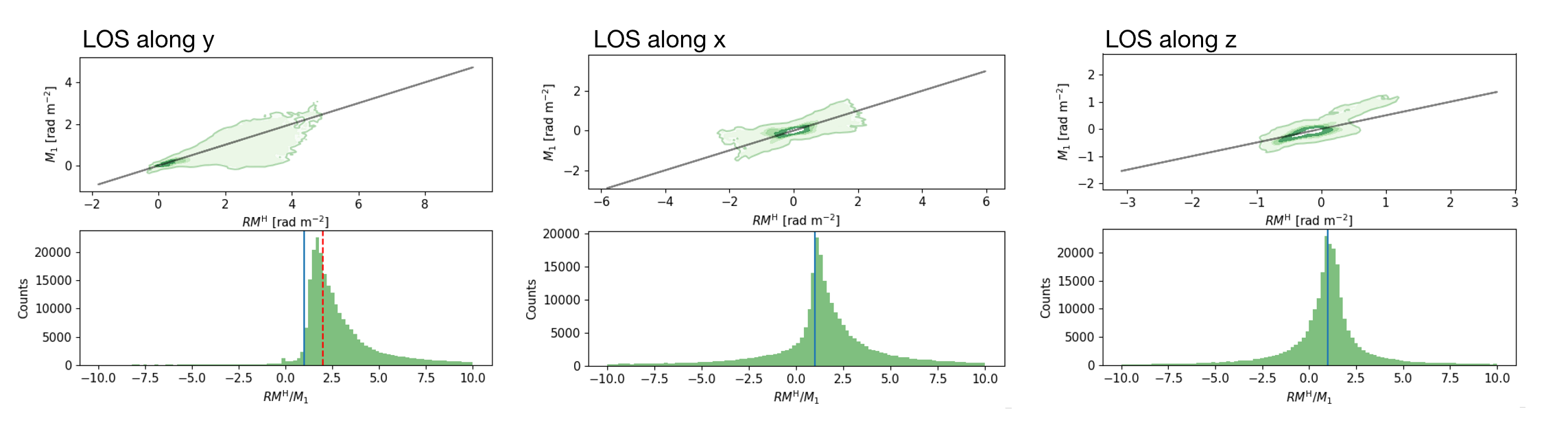}}
\caption{Correlation diagrams between $RM^{\rm H}$ and $M_1$ (top row) together with the histograms of their ratio (bottom row). In the top panels, the $M_1$ = $RM^{\rm H}/2$ line is shown in gray. In the bottom panels, the vertical blue and red lines correspond to ratios of 1 and 2, respectively. Three different lines of sight of case A are shown.}
\label{fig:rmm1}
\end{center}
\end{figure*}

This can be better seen on the left of Fig.~\ref{fig:spectra}, where the 2D angular power spectra, $p(k)$, of Stokes $I$ and $PI$ are shown. We notice that $p(k)$ of Stokes $I$ has a bump at about $k \sim 0.03$. This is likely due to the prominent filamentary structure in the middle that has a typical width of $\sim$30 px. Both power spectra are affected by the beam at the largest $k$ values. The power spectrum of Stokes $I$ is steeper than that of $PI$ with power-law indices in $k$ space of $-3.9$ and $-2.5$, respectively. We checked that such difference found at 150 MHz is not observed at higher frequency ($\sim$20 GHz), where Faraday rotation is negligible. 

The effect of differential Faraday rotation in polarization is more impressive when looking at Faraday tomographic data of $PI$ as a function of $\phi$. For case A (integrated along the $y$ axis) Fig.~\ref{fig:tomo} illustrates three slices of $PI$ in units of mJy PSF$^{-1}$ RMSF$^{-1}$ at given $\phi$ of 0, 1, and 2 rad m$^{-2}$, respectively. The apparent resemblance of our mock observations of Faraday tomography with actual data of polarized diffuse emission detected with LOFAR is striking \citep[e.g.,][]{Jelic2014,vanEck2017,Turic2021}. Regions of high $PI$ emission show patchy and filamentary structures close to linear, narrow, and depolarized features that highly resemble the so-called depolarization canals found in real data \citep[e.g.,][]{Haverkorn2003a,Jelic2018}. A detailed analysis of these features in our simulations will be the subject of future work.    
Here, we limit ourselves to exploring the statistics of the Faraday tomographic cubes using the Faraday moments first introduced by \citet{Dickey2019} to study Galactic polarized emission above 300 MHz with the Galactic Magneto-Ionic Medium Survey. Using Eqs. 5, 6, and 7 presented in \citet{Dickey2019} we define the moments $M_0$, $M_1$, and $M_2$ that encode the total $PI$ in the Faraday tomographic cube, the mean-weighted $\phi$, and its corresponding variance, respectively. For the same simulations as those shown in Fig.~\ref{fig:tomo}, we present the moments in Fig.~\ref{fig:moments}. $M_0$ and $\sqrt{M_2}$ are strongly correlated to each other, while $M_1$ shows distinct patterns. As also recently discussed by \citet{Erceg2022} in the analysis of the LoTSS survey, the values of $\sqrt{M_2}$  trace complex lines of sight resulting from differential Faraday rotation.

The structure of $M_1$ should be a proxy of the $RM$ along the LOS. However, as can be seen by comparing the map of $M_1$ with that of $RM^{\rm H}$ (see Fig.~\ref{fig:rmtot}) differences arise. These differences are caused by differential Faraday rotation. In Fig.~\ref{fig:rmm1} we compare $M_1$ and $RM^{\rm H}$ for the projections along the coordinate axes of case A. We show both 2D (top row) and 1D histograms of their relative ratio ($RM^{\rm H}/M_1$, bottom row). The 2D histograms indicate a spread of $M_1$ values around the $M_1 \propto 0.5\, RM^{\rm H}$ relation, regardless of the integration axis. However, the histograms of the ratio reveal that a value of the peak greater than unity ($\sim$2) is only observed when the integration axis is along the mean direction of the $\vec{B}$ field.   

\subsection{Multiphase gas contribution to polarized intensity}\label{ssec:phasepi}

As just shown, differential Faraday rotation strongly affects the amount of $PI$ that can be detected at low radio frequencies. Since differential Faraday rotation depends on the multiphase structure of the intervening ISM, in this section we investigate what is the contribution of each gas phase (see Sect.~\ref{ssec:phases}) to $PI_{\phi}$. 

We addressed this question by studying the correlation between the morphological structure in the maps of $PI_{\phi}$ and each $N_{\rm H}$ map introduced in Sect.~\ref{ssec:phaserm}. We used the $V$ parameter from HOG to quantify the relative alignment between the local gradients of $PI_{\phi}$ with those of the total gas column density, $N_{\rm H}$, of each phase. As explained in detail in Appendix~\ref{app:rmsf}, because of the shape of the RMSF, we weighted the $V$ parameter from HOG with the ratio between the maximum value of $PI_{\phi}$ at a given slice and the maximum value of the full Faraday tomographic cube. As a null-test reference, we also studied the correlation between $PI_{\phi}$ with a map produced from a Gaussian random field, for which no morphological correlation is expected. We chose a Gaussian random field characterized by a power-law power spectrum with index of $-2.7$ so to introduce some multi-scale structure in the random map.   

\begin{figure*}[!h]
\begin{center}
\resizebox{\hsize}{!}{\includegraphics{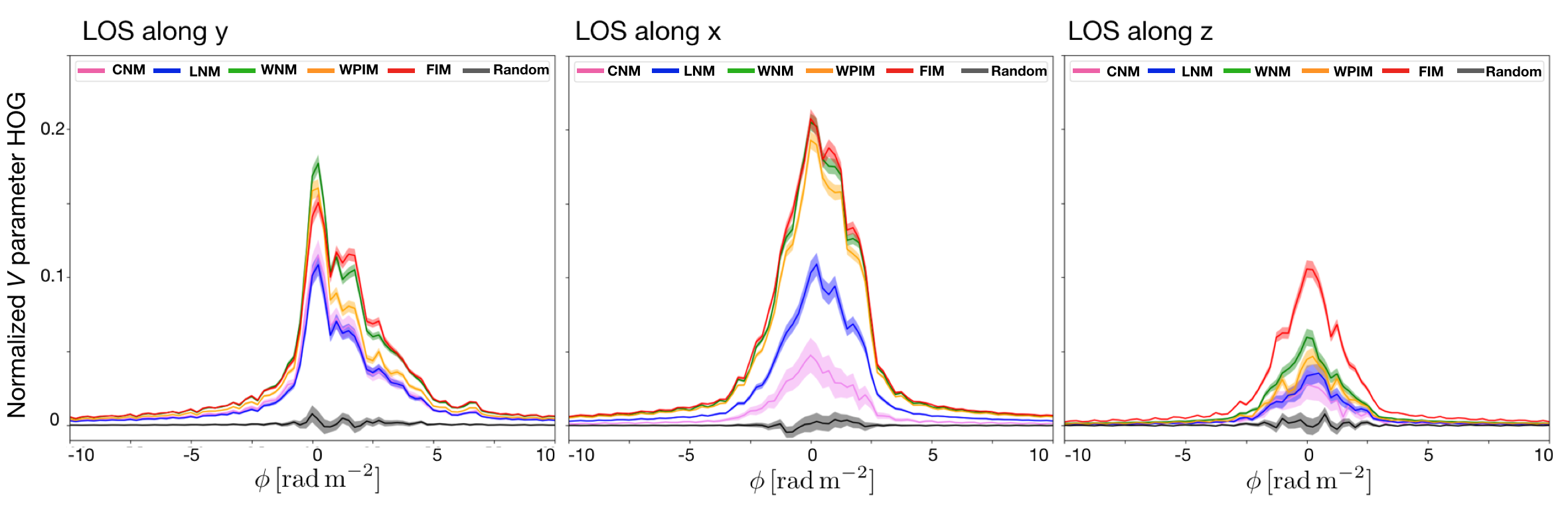}}
\caption{Correlation analysis between $PI_{\phi}$ and the $N_{\rm H}$ maps of each different gas phase (see the encapsulated legend) based on the projected Rayleigh statistics obtained with HOG. As a reference, the correlation with a random map is also shown in black. Three different LOS of case A are shown.}
\label{fig:hogsims}
\end{center}
\end{figure*}

\begin{figure}
\begin{center}
\resizebox{\hsize}{!}{\includegraphics{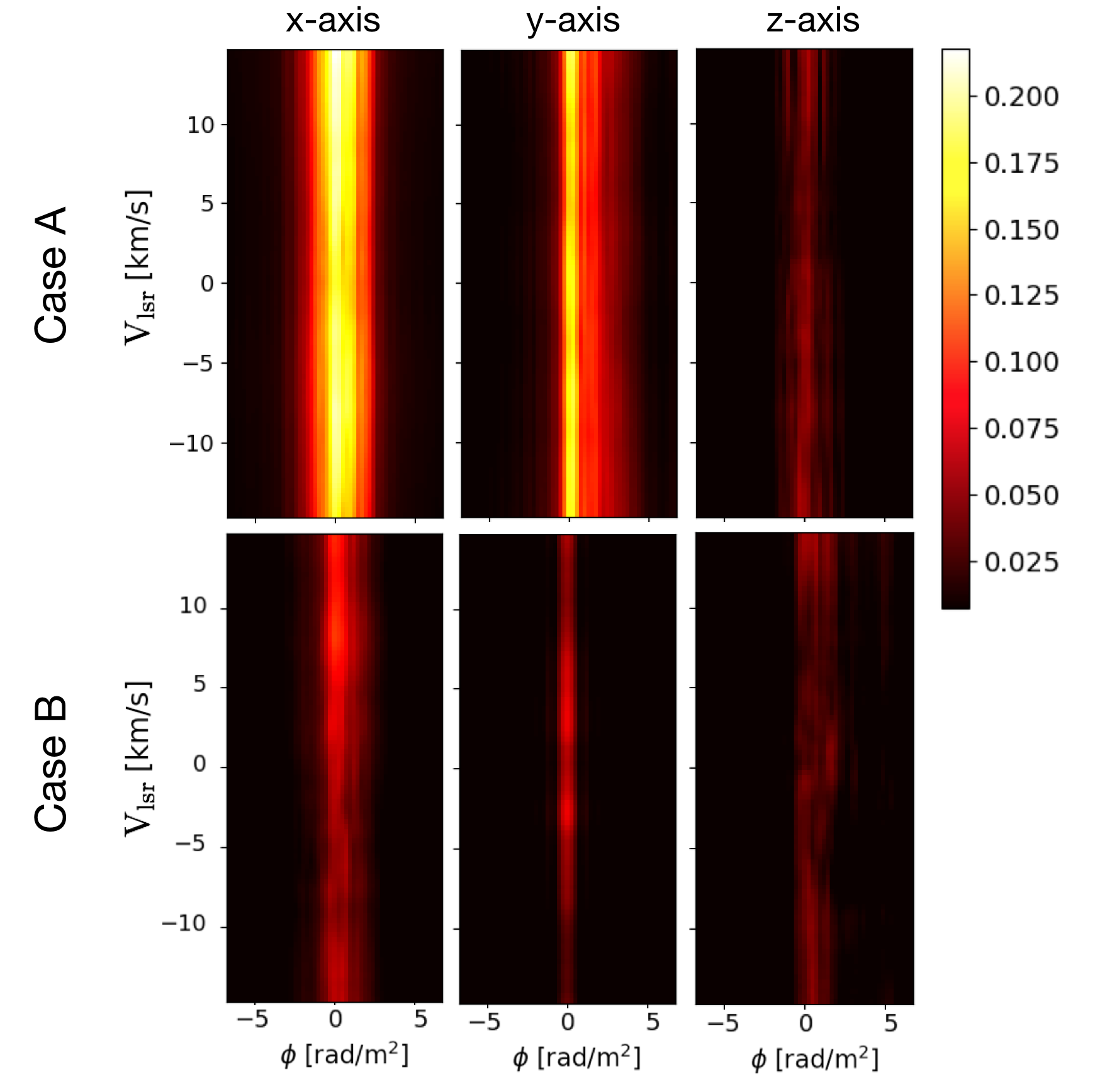}}
\caption{Normalized $V$ parameter from HOG between synthetic maps of $PI$ function of $\phi$ and of HI brightness temperature, $T_{\rm b}$, function of $V_{\rm lsr}$. The corresponding LOS integration axis and physical scenario are labeled. The color scale is the same for all panels.}
\label{fig:Vproj}
\end{center}
\end{figure}

In Fig.~\ref{fig:hogsims} we show the resulting $V$ parameter as a function of $\phi$ for all gas phases and integration axes of case A. Most of $PI$ appears correlated with the structure of $N_{\rm H}$ at low absolute values of $\phi$. The correlation with the different gas phases significantly depends on the integration axis. Generally speaking, all phases show morphological correlation with $PI$ compared to the random test, although the correlation is the strongest for WNM, WPIM, and FIM. Interestingly, the WNM and WPIM are those that correlate the most for the integration along the $y$ axis. This is not true when the LOS is along the $z$ axis (shell collision seen face on). In this case all phases appear poorly correlated with $PI$ except for the FIM. 

The synthetic $PI$ that survives the differential Faraday rotation in our simulations shows a complex correlation among gas phases that strongly depends on the choice of the integration axis. In order to bridge our models with real observables, rather than with $N_{\rm H}$ maps, we also studied the morphological correlation between $PI_{\phi}$ and synthetic observations of brightness temperature, $T_{\rm b}$, of optically thin HI emission, as first presented in \citet{Bracco2020b} with real data. We used the publicly available code {\tt BT-21cm}\footnote{\url{http://github.com/BarbaraSiljeg/Brightness-temperature-of-21-cm-line-from-a-simulation}} that, given the simulated $n_{\rm H}$, $T$, and LOS velocity, produces estimates of $T_{\rm b}$, as a function of local-standard-of-rest velocity ($V_{\rm LSR}$), in units of K following standard radiative transfer of HI \citep[e.g.,][]{Spitzer1978,mamd2007}. 

We quantified the correlation between $PI_{\phi}$ and $T_{\rm b}(V_{\rm LSR})$ using HOG. We built maps of the normalized $V$ parameter, as explained above and in Appendix~\ref{app:rmsf}, for all cases and integration axes. These maps are shown in Fig.~\ref{fig:Vproj} as a function of $\phi$ and $V_{\rm LSR}$. 

The appearance of the $V$-parameter maps shows strong correlation between $PI_{\phi}$ and $T_{\rm b}(V_{\rm LSR})$ for low absolute values of $\phi$, as also expected from Fig.~\ref{fig:hogsims}. The dependence on $V_{\rm LSR}$ is not as well defined. This is due to the dominant contribution of WNM over LNM or CNM (see Table~\ref{tab:volfrac}). The WNM has wide spectroscopic lines \citep{Wolfire2003}, which give rise to bright elongated features in the $V$-parameter maps. Figure~\ref{fig:Vproj} shows that, at least for the WNM phase, we are able to reproduce the observed correlation between $PI$ and HI emission as reported in  \citet{Bracco2020b}. 

We notice that this correlation highly depends on both integration axis and on the physical scenario taken into account. As expected, the correlation between $PI$ and $T_{\rm b}$ is the strongest when the LOS integration occurs perpendicular both to the mean magnetic-field direction and to the shell-collision axis, as for case A with the LOS along the $x$ axis. In case B the correlation between $PI$ and $T_{\rm b}$ is always lower than case A. This is due to the different physical evolution of the phases. In particular, as also listed in Table~\ref{tab:volfrac}, the WNM phase in case B occupies almost half of the simulated volume compared to case A. In case B the two shells collide along the mean magnetic field so that gas can transition more rapidly from WNM to colder phases.        

\section{Discussion}\label{sec:discussion}

For the first time, our study showed how important it is to account for the mutual interaction among ISM gas phases in order to model diffuse polarization detected at low radio frequencies. Nevertheless, we acknowledge the limitations of the analytical, steady-state, approach we used to define $n_{e}$ and to identify the five gas phases listed in Table~\ref{tab:phases}. We already commented on the strong parametric dependence of $n_{e}$ on the value of $\zeta$ (see Sect.~\ref{ssec:ne}), which requires more detailed studies on CR propagation models across the multiphase ISM \citep[e.g.,][]{Padovani2018,Kempski2021}. Moreover, time-dependent chemistry should be considered in order to properly account for the ionization state of warm and cold gas phases \citep[e.g.,][]{DeAvillez2020}, which inevitably affects the process of differential Faraday rotation \citep{Rappaz2022}. However, the error we make by applying Eq.~(\ref{eq:ne}) only underestimates the true amount of ionized gas \citep{deAvillez2012}. This means that in this work we provided lower limits to the Faraday rotation from the multiphase and ionized ISM that has an impact on the observed synchrotron polarization at low radio frequencies. Despite the caveats behind the estimate of $n_e$, we stress that our approach is a step forward in modelling low-frequency synchrotron polarization affected by Faraday rotation based on MHD simulations. To our knowledge, this work is the first attempt to study synthetic observations at low frequency using simulations that include distinct gas phases ranging over several orders of magnitude in $T$ and $n_{\rm H}$.

Previous works already made the effort of studying numerically the complexity of Faraday rotation and tomography. However they limited their studies to isothermal cases, expressing $n_e$ as a constant fraction of $n_{\rm H}$ \citep[e.g., ][]{Basu2019,Seta2021}. This is possibly the reason why in \citet{Seta2021} the authors could not generally apply Eq.~(\ref{eq:avbpar}) to derive $\langle B_{\parallel} \rangle_{\rm sim}$ from $\langle B_{\parallel} \rangle_{\rm pul}$. As they considered an isothermal ideal MHD simulation, a much tighter correlation between $\vec{B}$ and $n_{e}$ could be seen, introducing a possible bias on the weighting of $\vec{B}$ along the LOS. In our models, on the contrary, as detailed in Sect.~\ref{ssec:rmneb}, $\vec{B}$ and $n_{e}$ are not correlated, validating Eq.~(\ref{eq:avbpar}).     

Compared to the works mentioned above, we produced more realistic models in terms of the properties of the multiphase gas, despite the very specific choice of our modelled physical scenario, namely that of two colliding super shells. Large-scale and shell-like polarization structures in the radio band have been extensively observed across tens of degrees in the sky, often referred to as loops \citep[e.g.,][]{berkhuijsen1971a,vidal15,planck16,Thomson2021,Panopoulou2021}. One of such structures, Loop III \citep[e.g.,][]{Spoelstra1972,Paseka1993}, represents the dominant structure in polarization within the largest mosaic of the LoTSS survey presently done and presented in \citet{Erceg2022}.
These loops are likely the result of multiple supernova explosions as those simulated by \citet{Ntormousi2017} and considered in this work. We believe that the choice of our case study is well motivated by observational evidence, although we recognize that our results and conclusions cannot be easily generalized to any MHD process in the ISM. 

In Sect.~\ref{ssec:phaserm} we addressed the key question about which gas phase might be the most relevant for the observed values of $RM^{\rm H}$. It is important to stress that the answer to this fundamental question is highly dependent on the physical scenario (cases A and B) and on the LOS integration axis. It is difficult to provide one simple rule of thumb based on our study, which means that in a general context, special care should be applied to interpreting $RM$ observations. Broadly speaking, the warm, partially or fully ionized phases are those that dominate $RM$ over the coldest and most neutral ones. However, we cannot exclude the possibility that this results from the very low volume filling fraction of CNM and LNM in the simulations (see Table~\ref{tab:volfrac}). We also found a similar result when considering the gas-phase contribution to $PI$ in Sect.~\ref{ssec:phasepi}. Interestingly, the contribution from WNM and WPIM is never negligible compared to that of FIM. This supports the idea, already proposed by \citet{Heiles2012}, that Faraday rotation and low-frequency polarization, rather than recombination lines like H$_{\alpha}$, could be a powerful probe of partially ionized gas, which presently challenges our understanding of structure formation in the local ISM \citep[e.g.,][]{Jenkins2013, Gry2017}. Moreover, the role of WNM is also highlighted by the correlation found between $PI$ and $T_{\rm b}$ in Fig.~\ref{fig:Vproj}. This result is reminiscent of the observational analysis of the LOFAR data done by \citet{Bracco2020b}, where $PI$ was found morphologically correlated with $T_{\rm b}$ of HI data from the Effelsberg telescope \citep{Winkel2016}. 

The choice of the LOS integration axis, however, has a huge impact on our results, particularly related to the anisotropy introduced by the mean-magnetic field direction in the simulations. We notice that several observables may give indication about this main source of anisotropy. First of all, the correlation between $PI$ and $T_{\rm b}$ is the strongest when the LOS is perpendicular to the mean-magnetic field direction. This could be the case, as suggested by \citet{Zaroubi2015} and \citet{Jelic2018}, for the observed correlation found between LOFAR polarization and tracers of neutral ISM in the surroundings of the 3C 196 field \citep{Bracco2020b,Turic2021}. We must comment, however, on the lack of correlation between the simulated $PI$ and the cold phases (CNM and LNM) in contrast to what was reported in \citet{Bracco2020b}, and previously in \citet{vanEck2017} using data of interstellar dust extinction. The authors claimed that most of the correlation between the neutral medium (both probed by HI and dust) and LOFAR diffuse polarization was coming from CNM. This discrepancy between observations and simulations remains an open issue. Most likely, because of the very low fraction of CNM in these simulations, we are not able to quantitatively address the role of CNM. 
More work with simulations is needed to solve this inconsistency. 

The anisotropy related to the mean magnetic field direction is also observable using the $RM^{\rm H}/M_1$ ratio as a proxy. As shown in Fig.~\ref{fig:rmm1}, we suggest that a peak of the distribution of $RM^{\rm H}/M_1$ different than unity could be indicative of looking along the mean-magnetic field. However, this result is dependent upon our ability to define $RM^{\rm H}$ and $M_1$ for a common LOS volume in actual, low-frequency data. The slope of 2 in the correlation plot between $RM^{\rm H}$ and $M_1$ is in turn well understood in terms of differential Faraday rotation in Faraday thick structures \citep[][]{Burn1966,ordog19,Erceg2022}.  

The effect of differential Faraday rotation is also revealed by the structures of Stokes $I$ and $PI$ at 150 MHz presented in Sect.~\ref{ssec:tomo}. The 2D angular power spectra of the two maps are significantly different, with that of the $PI$ map being flatter. The small-scale structure in polarization introduced by differential Faraday rotation is a characteristic feature only observable at these low radio frequencies. We notice that real LOFAR data have the property of not showing any -- or mostly any -- Stokes $I$ counterpart of the diffuse polarized emission detected across all fields of view observed so far \citep[][]{Jelic2014,Jelic2015,vanEck2017,Turic2021}. 
As is known from previous Westerbork polarized observations \citep{Wieringa1993}, the missing short spacings in radio interferometers affect the large-scale emission of Stokes $I$ more severely than that of $PI$, where the small-scale polarized structure survives.

A more careful analysis with simulations, including realistic models for the instrumental characteristics of LOFAR \citep[for instance using {\tt OSKAR}\footnote{\url{https://github.com/OxfordSKA/OSKAR}} as in][]{Mort2017}, is beyond the scopes of this work and will be part of future studies.  

\section{Summary and conclusion}\label{sec:sum}

Faraday tomographic data below 200 MHz from the LOFAR telescope are challenging our understanding of the multiphase and magnetized ISM \citep[e.g.,][]{Jelic2014,Zaroubi2015,vanEck2017,Bracco2020b}. In this work we presented the first-ever analysis of synthetic data derived from MHD numerical simulations of Faraday tomography including multiphase ISM with temperatures and densities varying over more than four orders of magnitude.  

We produced mock observations of differential Faraday rotation of synchrotron polarized emission between 115 MHz and 170 MHz, reaching values of Faraday depth similar to those observed with LOFAR in the Galactic ISM between $-10$ and $+10$ rad m$^{-2}$. We used simulations of two colliding super shells produced by stellar feedback presented in \citet{Ntormousi2017}. 

The main results of our study are the following. Realistic MHD simulations reveal that the coexistence of gas phases (from fully ionized to cold neutral media, CNM) is key to interpreting data affected by differential Faraday rotation observed at low radio frequency. The multiphase ISM leaves its imprint both in the analysis of rotation measure data and of Faraday tomographic data. In the case of rotation measure, our analysis showed that most of its structure is related to the structure of the intervening magnetic field. However, the contribution of the electron density is not negligible. In particular, we found that the warm and partially ionized phases (WNM and WPIM) may represent a large contribution to the observed rotation measure. Similarly, we found that these phases also contribute to  most of the polarized intensity ($PI$) detected between 115 MHz and 170 MHz. 

All results strongly depend on the LOS integration axis and on the physical scenario under study. We explored two different cases, in which the super-shell collision axis is either perpendicular or parallel (cases A and B, respectively) to the mean-magnetic field direction in the simulations. Using synthetic spectroscopic observations of atomic hydrogen (HI), we found that the correlation between WNM and $PI$ is the strongest for case A, when the LOS is perpendicular to the mean-magnetic field direction and to the shell-collision axis. This result supports the interpretation already provided to explain the observational correlation found between LOFAR data and HI data toward the 3C 196 field \citep{Kalberla2016,Bracco2020b}.
On the other hand, regardless of the LOS, we found that our simulations always validate (within 1-$\sigma$ deviation) the phenomenological derivation of the line-of-sight average magnetic-field strength as proposed in studies of Galactic pulsars \citep[e.g.,][]{Sobey2019}.  

One open issue that arises from our work is that our simulations do not show a strong relation between $PI$ and CNM structures, while the analysis of real observations hints at the possibility of one \citep{vanEck2017,Bracco2020b}. We notice that this inconsistency may be related to the low volume fraction of CNM in our simulations (a few \%). However, we cannot exclude that other physical processes, not captured by the assumptions we made to model the ionization state of the ISM, or not related to the specific super-shell scenario, may be at play to justify the CNM issue. 

As discussed in this pioneering exploratory study, additional work on MHD simulations is needed in order to investigate more carefully the complexity of the multiphase and magnetized ISM and its imprint on low-frequency polarization. Such an effort will be crucial for Galactic magnetism studies at low radio frequencies and interpreting data from future large-scale surveys both from LOFAR in the north \citep{Shimwell2017,Shimwell2022} and from the Square Kilometre Array and its precursors in the south \citep{Dewdney2009}.

\begin{acknowledgements}
We are grateful to the anonymous referee for his/her comments. We are thankful to M.-A. Miville-Desch{\^e}nes, P. Hennebelle, F. Boulanger, and J. D. Soler for useful discussions. AB acknowledges the support from the European Union’s Horizon 2020 research and innovation program under the Marie Skłodowska-Curie Grant agreement No. 843008 (MUSICA). AB is also deeply grateful to Delphine for maternal patience and profound inspiration:"{\it Qu\'e he sacado con el lirio, que plantamos en el patio, no era uno el que plantaba, eran dos enamorados, ay ay ay}". AB welcomes Elio to this turbulent world. VJ, LT and AE acknowledge support by the Croatian Science Foundation for the project IP-2018-01-2889 (LowFreqCRO). VJ and LT also acknowledge support by the Croatian Science Foundation for the project DOK-2018-09-9169.
EN is supported by the ERC Grant ”Interstellar” (Grant agreement 740120).
The authors acknowledge Interstellar Institute's program "The Grand Cascade" and the Paris-Saclay University's Institut Pascal for hosting discussions that nourished the development of the ideas behind this work. This research made use of Astropy,\footnote{\url{http://www.astropy.org}} a community-developed core Python package for Astronomy \citep{astropy2013, astropy2018}.

\end{acknowledgements}

\bibliographystyle{aa}
\bibliography{main.bbl}
\appendix 
\onecolumn
\section{Supplementary figures}\label{app:supp}

In this Appendix we present figures that support some of the results presented in Sect.~\ref{ssec:rmneb} and Sect.~\ref{ssec:phaserm}.

\begin{figure*}[!h]
\begin{center}
\resizebox{1\hsize}{!}{\includegraphics{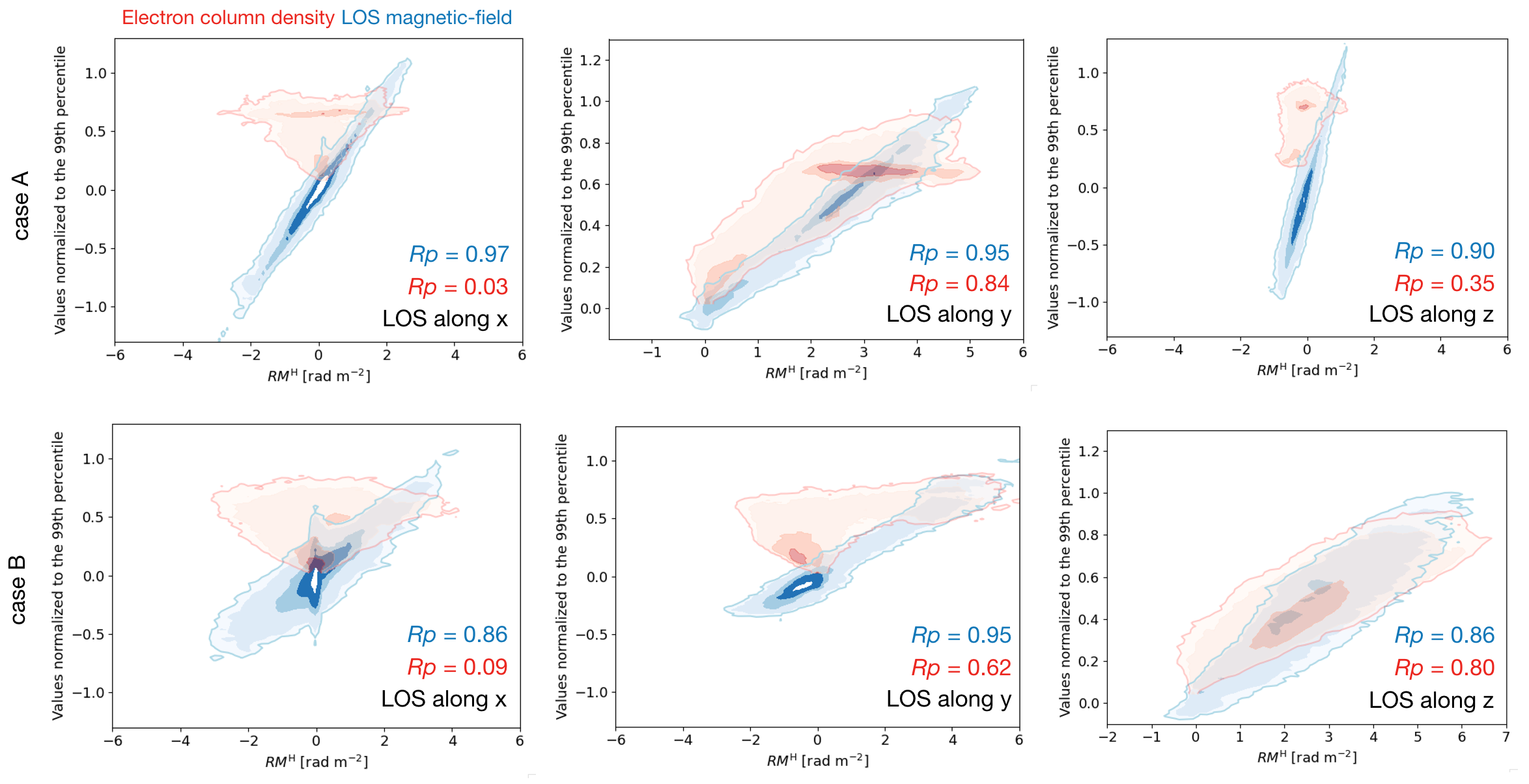}}
\caption{Same as in Fig.~\ref{fig:rmbnecor} but for lines of sight (LOS) along the $x$ and $z$ axes for case A (top row) and for all LOS axes for case B (bottom row).}
\label{fig:rmbna1}
\end{center}
\end{figure*}



\begin{figure*}[!h]
\begin{center}
\resizebox{1\hsize}{!}{\includegraphics{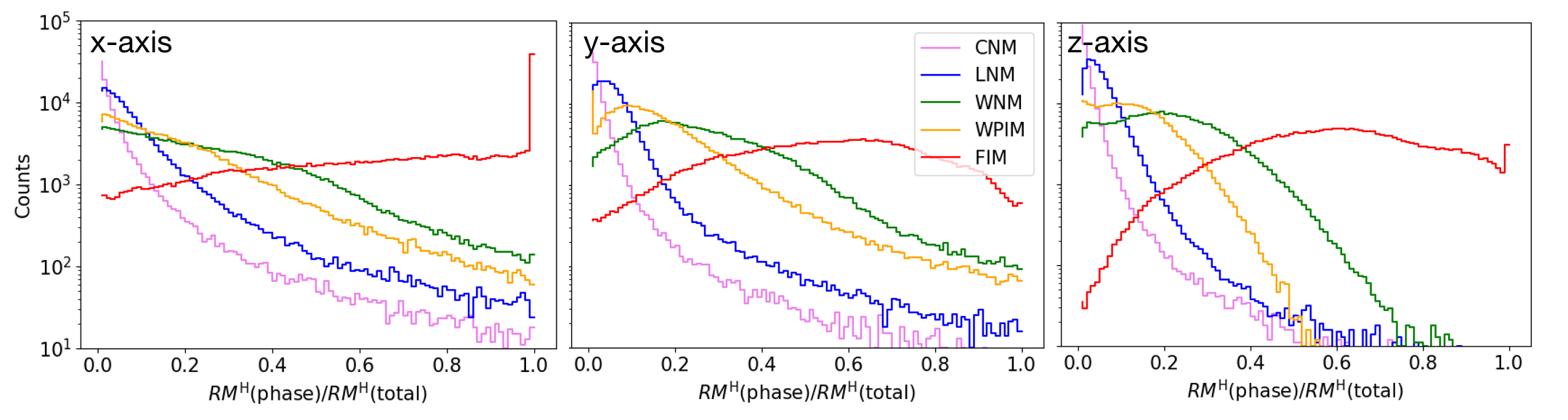}}
\caption{Same as in Fig.~\ref{fig:rmrel1} but for case B. Colors are defined in the central panel.}
\label{fig:rmrel2}
\end{center}
\end{figure*}

\twocolumn
\section{Rotation measure spread function and HOG}\label{app:rmsf}

As mentioned in Sect.~\ref{ssec:phasepi}, the use of HOG with maps of $PI_{\phi}$ must take into account the shape of the RMSF resulting from low-frequency Faraday tomography.

\begin{figure}[!h]
\begin{center}
\resizebox{0.9\hsize}{!}{\includegraphics{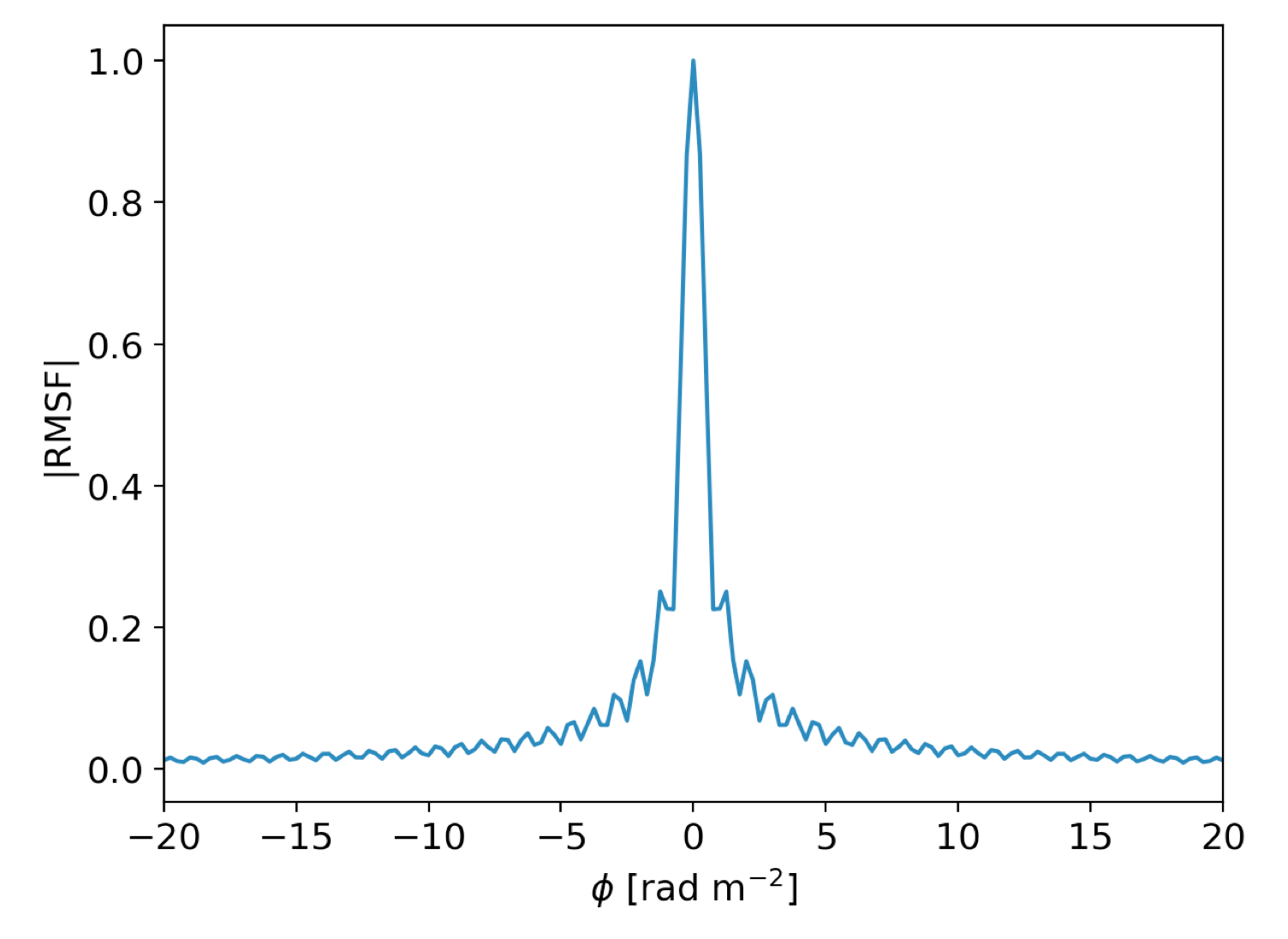}}
\caption{Rotation measure spread function used in this work to perform Faraday tomography.}
\label{fig:rmsf}
\end{center}
\end{figure}

The RMSF of the synthetic Faraday spectra at LOFAR frequencies has side lobes (see Fig.~\ref{fig:rmsf}), which produce leakage from polarized intensity at a given $\phi$ over the full Faraday spectrum. This means that the structure of $PI_{\phi}$ reproduces itself at the peak of each side lobe in Faraday space. Thus, since in our analysis we did not introduce polarization noise, which, if large enough, can hide the side-lobe leakage, we faced the problem of identifying the right $\phi$ values for which $PI_{\phi}$ would truly correlate with any of the $N_{\rm H}$ maps or $T_{\rm b}$ maps. 

\begin{figure}[!h]
\begin{center}
\resizebox{0.9\hsize}{!}{\includegraphics{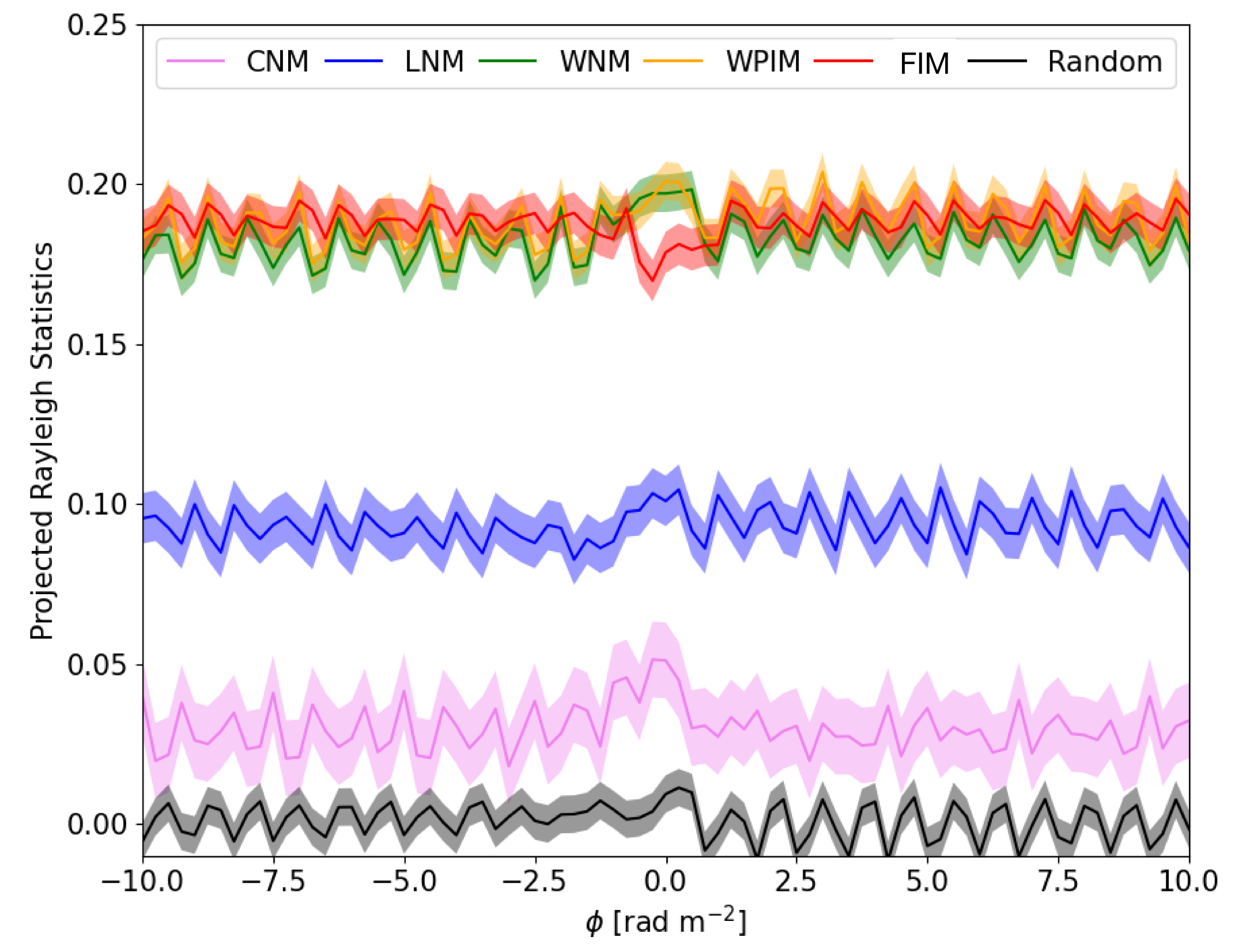}}
\caption{Same as in Fig.~\ref{fig:hogphase} without normalizing the $V$ parameter from HOG to the ratio between the maximum value of $PI_{\phi}$ at a given slice in the Faraday cube and the maximum value of the full Faraday tomographic cube.}
\label{fig:biasrmsf}
\end{center}
\end{figure}

As an example, in Fig.~\ref{fig:biasrmsf} we show how the same plot as the one presented in the middle panel of Fig.~\ref{fig:hogphase} would look like without weighting the $V$ parameter from HOG to the ratio between the maximum value of $PI_{\phi}$ at a given slice in the Faraday cube and the maximum value of the full Faraday cube.  

If on the one hand one could still identify the relative contribution of each phase to $PI_{\phi}$, on the other hand it would not be possible to distinguish the right range of $\phi$ that would correspond to the morphological alignment between $PI_{\phi}$ and the $N_{\rm H}$ maps of the different gas phases. 




\end{document}